\documentclass[twocolumn,twoside]{IEEEtran}
\usepackage{listings}
\usepackage{cite}
\usepackage{url}
\usepackage{xcolor}
\usepackage{graphicx}
\usepackage{epstopdf}
\usepackage[cmex10]{amsmath}
\usepackage{amsfonts}
\usepackage{amssymb}
\usepackage{amsthm}
\usepackage{mathrsfs}
\usepackage{verbatim}
\usepackage[noend]{algpseudocode}
\usepackage{algorithmicx,algorithm}
\usepackage{setspace}
\usepackage{bm}%
\usepackage{stfloats}
\usepackage{enumerate}
\usepackage{colortbl}
\usepackage{color}
\usepackage{array}
\theoremstyle{plain}

\newtheorem{lemma}[]{Lemma}
\newtheorem{corollary}{Corollary}

\newtheorem{Theorem}[]{Theorem}
\newtheorem{remark}{Remark}
\usepackage{subfigure}
\usepackage{multirow} 
\usepackage[caption=false,font=normalsize,labelfont=sf,textfont=sf]{subfig}
\begin{document}
        \title{Age-Threshold Slotted ALOHA for Optimizing Information Freshness in Mobile Networks}
	\author{Fangming~Zhao, Nikolaos Pappas, Chuan Ma, Xinghua Sun, Tony Q. S. Quek, and Howard H.~Yang}
	
	\author{
		Fangming~Zhao, 
		Nikolaos Pappas, \textit{Senior Member, IEEE},
		Chuan Ma, \textit{Member, IEEE}, \\
		Xinghua Sun, \textit{Member, IEEE}, 
      	Tony~Q.~S.~Quek, \textit{Fellow, IEEE},
		and  Howard~H.~Yang, \textit{Member, IEEE}
		
				
				
			
			
			
	}
 
	\maketitle
\begin{abstract}
    We optimize the Age of Information (AoI) in mobile networks using the age-threshold slotted ALOHA (TSA) protocol. The network comprises multiple source-destination pairs, where each source sends a sequence of status update packets to its destination over a shared spectrum. The TSA protocol stipulates that a source node must remain silent until its AoI reaches a predefined threshold, after which the node accesses the radio channel with a certain probability. Using stochastic geometry tools, we derive analytical expressions for the transmission success probability, mean peak AoI, and time-average AoI. Subsequently, we obtain closed-form expressions for the optimal update rate and age threshold that minimize the mean peak and time-average AoI, respectively. 
    In addition, we establish a scaling law for the mean peak AoI and time-average AoI in mobile networks, revealing that the optimal mean peak AoI and time-average AoI increase linearly with the deployment density. Notably, the growth rate of time-average AoI under TSA is half of that under SA. When considering the optimal mean peak AoI, the TSA protocol exhibits comparable performance to the traditional slotted ALOHA protocol. These findings conclusively affirm the advantage of TSA in reducing higher-order AoI, particularly in densely deployed networks.
\end{abstract}

	\section{Introduction}
 
Timeliness has emerged as a fundamental prerequisite for Internet-of-Things (IoT) services, encompassing various applications such as autonomous driving, smart factory operations, and intelligent healthcare. Acquiring timely updates from massively deployed sensors is crucial for making prompt and accurate decisions in these applications\cite{pappas2023age}. 

To assess the timeliness of delivered messages, the notion of Age of Information (AoI) has been proposed in \cite{AoIinfocom2012}\cite{6195689}, where the authors showed that minimizing AoI is fundamentally different from optimizing conventional metrics such as delay or throughput. 
As a result, plenty of research efforts have been devoted to ensuring timely information delivery over communication networks. 
This paper employs the age-threshold slotted ALOHA (TSA), a variant of the classic ALOHA protocol, to enhance AoI in random access networks. Specifically, TSA exploits the AoI information contained in the feedback messages to control each transmitter's channel access, thereby improving AoI. We study the minimization of AoI through the lens of the signal-to-interference-plus-noise ratio (SINR) model, by optimally adjusting the parameters of TSA according to the network configuration. 
The resultant scheme has low complexity and is particularly relevant to large-scale wireless systems. 
	
\subsection{Related Works}
In the early stages of AoI analysis for large-scale networks, many researchers adopted the traditional analytical framework of slotted ALOHA. For example, \cite{8006544} characterized and optimized the average AoI by optimal tuning the channel access probability. \cite{10114593} optimized the peak AoI using First-Come-First-Serve (FCFS) and Last-Come-First-Serve (LCFS) strategies, showing that with optimal packet arrival rate and channel access probability, the peak AoI scales linearly with the number of access nodes. 
\cite{9181539} considered a multi-channel ALOHA network, enhancing AoI performance via joint radio access control and resource allocation. \cite{TwoUserMA} explores the average AoI performance in a heterogeneous random access network where stochastic arrival nodes coexist with generate-at-will nodes. The function of irregular repetition slotted ALOHA (IRSA) on various AoI performance metrics has been demonstrated in \cite{IRSA}. The dynamic frame slotted ALOHA (FSA) has been studied in \cite{ICC2015}. 
These efforts have alleviated the performance bottlenecks of the AoI metric in large-scale networks. However, the primary objective behind the conventional ALOHA-like protocol was to mitigate channel contention, thereby enhancing transmission success probability or throughput. It was not explicitly designed to cater to the distinct attributes of the AoI metric. Hence, it becomes essential to explore random access protocols that closely align with the attributes of the AoI metric.

To design an age-aware random access network, it is essential to (a) mitigate network interference, (b) minimize the waiting time for information packets at the transmitters, and (c) ensure equalization of the update packet reception intervals. A straightforward strategy to address these needs is to prioritize nodes with lower timeliness for updates, allowing nodes with higher information freshness to remain silent temporarily. Based on this idea, several works \cite{ThresholdInfocomWS, yavascan2020analysis, ThresholdISIT,ThresholdWhittleIndex,MiSTA,TCOM2024AFSA} extended the Slotted ALOHA (SA) into an age-aware variant. Specifically, a threshold-based age-dependent random access protocol was proposed in \cite{ThresholdInfocomWS,yavascan2020analysis}, where each node accesses the channel only when its instantaneous AoI exceeds a predetermined threshold. A distributed transmission strategy was proposed in \cite{ThresholdISIT} based on the age gain, defined as the reduction of instantaneous AoI when packets are successfully delivered. An Index-Prioritized Random Access scheme was proposed in \cite{ThresholdWhittleIndex}, where nodes access the channel according to their indices that reflect the urgency of the update. Minislotted threshold ALOHA, designed to minimize the average AoI in the high payload scenario, is investigated in \cite{MiSTA}. The dynamic FSA with age gain threshold is considered in \cite{TCOM2024AFSA}.

\begin{table*}
\centering
\caption{A summary of the literature on the age-threshold aloha policy.}\label{tab:my_label}
\scalebox{0.7}{
\begin{tabular}{|c|c|c|c|c|c|c|c|c|c|c|}
\hline & \multicolumn{3}{|c|}{ Interference Analysis } & \multicolumn{2}{|c|}{ AoI Performance Analysis } & \multicolumn{2}{|c|}{ Optimal Configuration } & \multirow[t]{2}{*}{ Performance Limits } & \multirow[t]{2}{*}{ Scaling law } \\
\hline & Decoding model &  Spatial Attribute & Stable Region & Numerical & Closed-form & Numerical & Closed-form & & \\
\hline \cite{ThresholdInfocomWS} & Collison model &   & & &  \checkmark  &  \checkmark  & & & \\
\hline \cite{yavascan2020analysis} & Collison model &   &  \checkmark  & &  \checkmark  & &  \checkmark  & &  \checkmark  \\
\hline \cite{ThresholdISIT} & Collison model &   & & &  \checkmark  &  \checkmark  & & &  \checkmark  \\
\hline \cite{ThresholdWhittleIndex} & Collison model &   & & &  \checkmark  &  \checkmark  &  \checkmark  & & \\
\hline \cite{MiSTA} & Collison model &   &  \checkmark  & &  \checkmark  & &  \checkmark  & &  \checkmark  \\
\hline \cite{TCOM2024AFSA} & Collison model &   & &  \checkmark  & &\checkmark & & & \\
\hline \cite{yang2023analysis} & SINR model &    \checkmark  & & &  \checkmark  & & & & \\
\hline Proposed Work & SINR model &  \checkmark  &  \checkmark  &  & \checkmark  &   &\checkmark &   \checkmark  &  \checkmark  \\
\hline
\end{tabular}}
\end{table*}
The aforementioned research endeavors predominantly concentrate on the single-cell setting, where multiple nodes transmit data to a central access point. However, it is imperative to consider the ad-hoc network scenario for a more comprehensive understanding. Another noteworthy aspect is that the interference in the above references is modeled by the collision model\cite{TITCollisionModel}, wherein concurrent transmissions always lead to communication failures.
	Such a model neglects essential physical elements in wireless channels, such as fading, path loss, and spatial network topology, oversimplifying the co-channel interference effects.
	In contrast, the SINR model is preferable to the update packet verdict on the receiver end, as it can adequately account for these critical features in the ad-hoc network scenario.
	
	Consequently, a line of recent studies has been carried out, analyzing and optimizing AoI in wireless networks on the basis of the SINR model\cite{ZhongyiSG, mankar2020spatial, TimeEventAoI,  yangUnderstanding, LCFSyang, SGThroughputAoI ,AoIopSun, VTC, energyAoItradeoff, FSA,TMCSG, TMClocal,TMCpowercontrol}. 
	Specifically, the upper and lower bounds of the time-average AoI in Poisson bipolar networks are provided in \cite{ZhongyiSG} based on the favorable/dominant system argument, and the upper bound is improved by reconstruction of the dominant system in \cite{mankar2020spatial}.  
	The authors in \cite{TimeEventAoI} derived the average peak AoI with different packet arrival patterns, i.e., the time-triggered and event-triggered traffic for large-scale wireless networks, while the authors in \cite{yangUnderstanding} and \cite{LCFSyang} analyze the AoI statistics under different buffer sizes and package management disciplines. 
	The authors in \cite{SGThroughputAoI} present a comprehensive study on the interplay between throughput and AoI in a cellular-based IoT network. 
	Explicit optimal expression of peak AoI of slotted ALOHA, and its optimal parameters are derived in \cite{AoIopSun}, the distinctions and similarities between the optimal parameters for the peak AoI and average AoI are extensively discussed in \cite{VTC}. The energy-AoI tradeoff is further explored in \cite{energyAoItradeoff}.  
	The authors in \cite{FSA} examine the AoI performance of random access networks under frame-slotted ALOHA protocol. 
	Locally adaptive strategies for optimizing AoI have been studied in \cite{TMCSG}, \cite{TMClocal}, and \cite{TMCpowercontrol}, where the channel access probability, update rate, and transmit power at each node are adjusted according to the local communication environment to minimize AoI. 
	
	The existing results primarily pertain to the conventional slotted ALOHA protocol, whilst
	the effects of the age threshold remain unexplored. 
	Although a recent work \cite{yang2023analysis} has investigated the effect of TSA on AoI performance in static Poisson bipolar networks, the analysis involves complicated expressions of the SINR meta distribution, prohibiting it from optimizing TSA to reap the full potential of this protocol. In this paper, we consider the AoI optimization under TSA in a high mobility case, with which we can establish tractable analysis and obtain closed-form expression of the optimal network configuration, AoI performance limits, and scaling law that provide insights into network designs.

	\color{black}
	\subsection{Contributions}
	The main contributions of this paper are summarized below.
	\begin{itemize}
		\item We develop a mathematical framework for analyzing the impact of the TSA protocol on the AoI performance of random access networks. 
		We establish a fixed-point equation to characterize the transmission success probability, encompassing key features such as the update rate, age threshold, channel fading, deployment density, and co-channel interference.
		\item We derive analytical expressions of the mean peak AoI and time-average AoI over the typical link. 
        Our analysis culminates in an optimization strategy to minimize these AoI metrics, achieved by a joint optimization of the update rate and age threshold. 
        Specifically, we investigate the performance limits of the mean peak AoI under TSA protocol and identify the corresponding parameter pairs that enable its minimization. We also propose an alternative iterative algorithm to achieve the optimal parameter configuration that minimizes the time-average AoI. Additionally, we derive closed-form approximations for these optimal parameters, which are asymptotically accurate. 
		
		\item We establish a scaling law of AoI in random access networks. Notably, it reveals that although the optimal mean peak AoI and time-average AoI increase linearly with the deployment density, the growth rate of time-average AoI under TSA is only half of that under SA.

        \item Considering the inherent instability of ALOHA-like networks, especially under conditions of high interference levels and traffic load, we propose parameter tuning strategies for both mean peak AoI and time-average AoI in the presence of a bistable region, aiming to bolster the robustness of the TSA protocol network.
	\end{itemize}
    To better highlight the technical contributions of this paper, we summarize existing literature that dealt with age-threshold policy in Table~\ref{tab:my_label}.
	\begin{table*}[htbp]
		\renewcommand\arraystretch{1.3}
		\caption{key notations}
		\begin{center}
			\begin{tabular}{|c|c||c|c|}
				\hline
            \textbf{Notations}&\textbf{Definition}&\textbf{Notations}&\textbf{Definition} \\
				\hline
				$\lambda$ & Spatial deployment density & $\hat{\Delta}_0$ & Mean peak AoI of typical link\\
				\hline
				$r$ &  Transceiver distance	&$\bar{\Delta}_0$  &Time-average AoI of typical link\\
				\hline
				$\eta$ & Update rate&$\hat{\eta}^{*}_{\vert A}$ & Optimal update rate (Fixed $A$, mean peak AoI) \\
				\hline
				$A$ & Age-threshold 	&$\hat{A}^{*}_{\vert \eta}$&  Optimal age-threshold (Fixed $\eta$, mean peak AoI)\\
				\hline
				$\alpha$ & Path-loss fading coefficient &$(\hat{A}^{*},\hat{\eta}^{*})$ & Optimal age-threshold $\&$ update rate (mean peak AoI)\\
				\hline
				$\rho$ & SNR & $\bar{\eta}^{*}_{\vert A}$ & Optimal update rate (Fixed $A$, time-average AoI) \\
				\hline
				$\theta$ & Decoding SINR threshold & $\bar{A}^{*}_{\vert \eta}$ & Optimal update rate (Fixed $\eta$, time-average AoI) \\
				\hline
				$p_\mathrm{s}$ & Transmission success probability  &$(\bar{A}^{*},\bar{\eta}^{*})$ & Optimal age-threshold $\&$ update rate (time-average AoI)\\
                \hline
                $p_{\mathrm{s}}^{L}$ & $p_\mathrm{s}$ in low efficiency state & $\hat{\Delta}_0^{*}$  & Optimal mean peak AoI  \\
                \hline
                $p_{\mathrm{s}}^{H}$ & $p_\mathrm{s}$ in high efficiency state & $\bar{\Delta}_0^{*}$  & Optimal time-average AoI \\
				\hline
			\end{tabular}
			\label{tab:KEYNOTATIONS}
		\end{center}
	\end{table*}

\section{System Model}
\subsection{Spatial Configuration}
We consider a random access network comprised of source-destination pairs. 
The source nodes are scattered according to a homogeneous Poisson point process (HPPP) $\Phi_{ \mathrm{s} } = \{ X_i \}_{i=1}^\infty$ of density $\lambda$. 
Each source is paired with a destination in the distance $r$ and oriented in a random direction. 
According to the displacement theorem\cite{SGbookBac,SGbookMartin}, locations of the destination nodes constitute an independent HPPP $\Phi_{ \mathrm{d} }$ of the same density.
We add a receiver at the origin to the point process $\Phi_{\mathrm{d}}$. 
We also add its tagged transmitter, denoted by $X_0$, to the point process $\Phi_{ \mathrm{s} }$. 
We refer to this link pair as the \textit{typical} one. {\footnote{This setting is a large-scale analog to the classical model of Random Networks \cite{TITcapacity}, where the fixed distance between each transmitter-receiver pair represents the average value. It is noteworthy that the analysis developed in this article can be extended to investigate the AoI performance in networks with centralized infrastructures where multiple sources can associate with a common destination and are located at random distances from the destinations \cite{MartinRandomDistance,MartinReyleignDistance}.}}
Based on Slivnyak's theorem\cite{SGbookBac,SGbookMartin}, we can concentrate on analyzing the performance of the typical link. A sequence of status information is generated at each source, encapsulated into information packets, and transmitted over a shared spectrum. 
When a source node sends out information packets, it transmits at a fixed power. 
We assume the channel between any pair of nodes is affected by the Rayleigh fading with a unit mean, which varies independently across time slots, and path loss that follows the power law. We also assume the received signal is subjected to white Gaussian thermal noise.
The main notations used throughout this article are summarized in Table \ref{tab:KEYNOTATIONS}.
	
\subsection{Temporal Pattern}
In this network, time is segmented into equal-length intervals, in which each slot is the minimum duration required to transmit an information packet (and receive feedback from the destination).
We assume the network is synchronized\cite{TWCUplinkSyn,IoTJUplinkSyn}.
Each source adopts the \textit{generate-at-will} policy \cite{SunyinTIT} for the status updating.
Particularly, if a node decides to transmit, it will generate a new sample at the beginning of the time slot and send the
information packet to the destination immediately. 
By the end of the same time slot, the packet is successfully decoded if the received SINR exceeds a decoding threshold; otherwise, the
transmission fails, and in that case, the packet is dropped. The delivery of packets, therefore, incurs
a delay of one time slot. Additionally, the positions of the links vary according to the \textit{high mobility model} \cite{AoIopSun,MartinHM}, i.e., spatial locations of the link pairs are (re)generated independently in each time slot according to the same Poisson bipolar network model. \footnote{By leveraging the framework of \cite{MartinMoveLocaldelay}, one can extend the results obtained in this paper to investigate AoI in a mobile network with spatial correlations over time.} 
	
\subsection{Transmission Protocol}
We employ the TSA protocol for status updating. 
Specifically, every source node stays inactive until its AoI reaches a predefined threshold, denoted by $A$, upon which the source node turns on the status updating mode: it generates a fresh sample with probability $\eta$ at the beginning of the subsequent time slot, encapsulates that information into a packet, and immediately sends it to the destination. 
If the received SINR surpasses a decoding threshold $\theta$, the transmission succeeds; the receiver then feeds back an ACK to the source, and the age is reset to one.\footnote{In this paper, we make the commonly adopted assumption that the ACK feedback is error-free \cite{MartinHM,yavascan2020analysis,ThresholdISIT}. It is, nonetheless, noteworthy that the analysis can be extended to account for the feedback error of ACK messages.} 
Otherwise, the source will generate a new sample in the next time slot with probability $\eta$. 
This probability is usually referred to as the \textit{update rate}.
	
\subsection{Performance Metric}
\begin{figure}		\includegraphics[width=8.6cm,height=4.4cm]{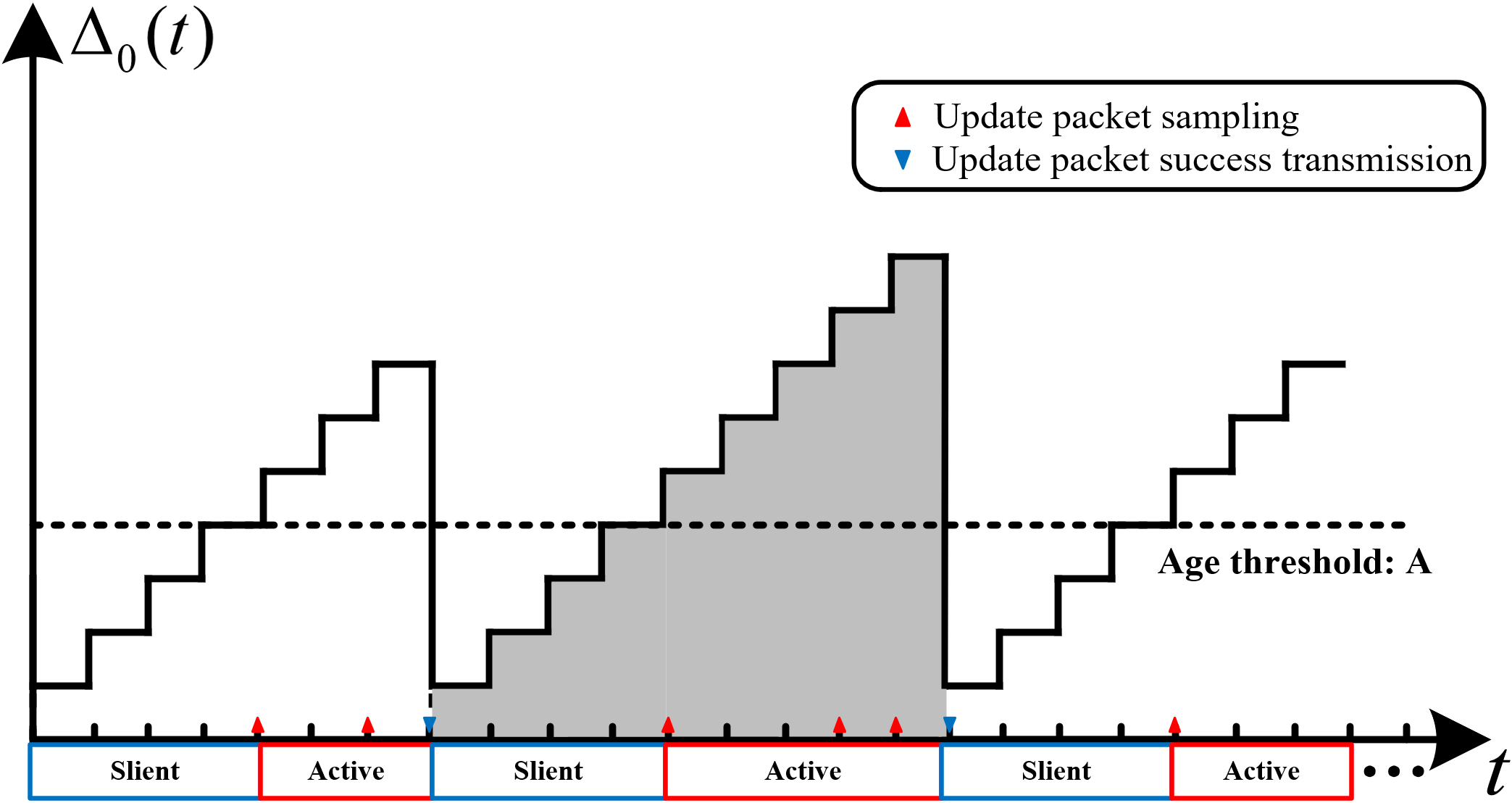}\\
		\caption{An illustration of the evolution of the age of information over
 the link under consideration in the discrete-time model.}
		\label{fig:AoI}
  \vspace{-0.5cm}
	\end{figure}

We put the main focus on the notion of AoI in this paper. 
As depicted in Fig.~\ref{fig:AoI}, AoI grows linearly with time in the absence of new
updates at the destination, and it drops to the time elapsed since the generation of the latest packet received. 
Formally, the age evolution process over the typical link can be written as follows \cite{AoIinfocom2012,SunyinTIT}
\begin{equation}
\Delta_0(t) = t- G_0(t),
\end{equation}
where $G_0(t)$ denotes the generation time of the latest packet successfully delivered over this link.
In this paper, we use the mean peak AoI and time-average AoI over the typical link as our performance metrics, defined respectively as \cite{yangUnderstanding} 
\begin{align}
    \hat{\Delta}_0 = \lim_{ N \rightarrow \infty } \frac{1}{N} \sum_{n=1}^N \Delta_0( T_0(n))
\end{align}
where $T_0(n)$ denotes the time slot at which the $n$-th packet is successfully delivered, and  
\begin{equation}
\bar{\Delta}_0 = \lim_{T\to \infty} \frac{1}{T}\sum_{t=1}^{T} \Delta_0(t).
\end{equation}
Since the point processes formed by the TSA protocol still exhibit ergodicity and stationarity, the statistical characteristics of AoI performance for the typical link are identical to those of the other links.

\section{AoI Analysis}
\subsection{SINR at the Typical Receiver}
Since all the source nodes use the same spectrum for packet delivery, each transmission would be affected by others owing to co-channel interference. As such, at the typical destination, the received SINR at time slot $t$ is\cite{UnifiedSINRyang}
\begin{equation}
    \mathrm{SINR}_0(t) = \frac{h_{00}(t) r^{-\alpha}}{\sum_{i\neq 0}h_{i0}(t)v_i(t)||X_i||^{-\alpha}+\rho^{-1}},
\end{equation}
where $h_{i0}(t) \sim \exp(1)$ represents the small-scale fading between transmitter $i$ and the typical receiver, $\alpha$ is the path-loss exponent, $\Vert \cdot \Vert$ stands for the Euclidean norm, $\rho$ denotes the signal-to-noise ratio (SNR) between the transmit power of each source node and the noise power, and $v_i(t)$ is a binary function, where $v_i(t)=1$ indicates that transmitter $i$ initiates a packet transmission at time slot $t$, and $v_i(t)=0$ otherwise. 

	\subsection{Transmission Success Probability}
	According to the transmission policy, an information packet is successfully delivered if the received SINR exceeds a decoding threshold $\theta$. Therefore, the probability of successfully transmitting a packet from the typical source node can be written as \cite{SGbookMartin}
	\begin{equation} \label{eq:define_p}
		p_{\mathrm{s}}(t) = \mathbb{P}\left(\mathrm{SINR}_{0}(t)>\theta\right).
	\end{equation}
	Under the high mobility random walk model, the received $\text{SINR}_i(t)$ of each transmitter $i$, $i \in \mathbb{N}$, is independent and identically distributed (i.i.d.) across time $t$. 
	By symmetry, the transmission success probability is also identical across the transmitters.
	As such, we drop the time index from $p_{\mathrm{s}}(t)$ in the sequel.
 The following theorem provides a formal characterization of this quantity. 
	\begin{Theorem} \label{lemma:define:p}
		The transmission success probability under TSA is given by the following fixed-point equation\begin{equation}\label{eq:define:p}
			p_{\mathrm{s}} = \exp\left(-\frac{ \lambda c \eta r^2 }{1+A\eta p_{\mathrm{s}}} -\frac{\theta r^{\alpha}}{\rho}\right),
		\end{equation}
  in which 
  	\begin{equation}
		c = \pi \theta^{\frac{2}{\alpha}}\Gamma\left(1-\frac{2}{\alpha}\right)\Gamma\left(1+\frac{2}{\alpha}\right),
	\end{equation}
 where $\Gamma(\cdot)$ denotes the Gamma function \cite{7345601}.
	\end{Theorem}
	\begin{IEEEproof}
	Please see Appendix \ref{proofTheorem1}. 
\end{IEEEproof}
The expression in \eqref{eq:define:p} accounts for the impact of several key factors, including the update rate, age threshold, path loss, and deployment density, on the transmission success probability. 
As opposed to other variants of SA, e.g., the FSA \cite{FSA}, the transmission success probability under TSA is given in the form of a fixed-point equation, which steams from the spatial-temporal interactions amongst the source nodes across the network.

The parameter $c$ in \eqref{eq:define:p} reflects the level of \textit{spatial contention} \cite{localDelayMartin}, measuring the network's capability of spatial reuse by quantifying how fast the transmission success probability deteriorates when the interferers' density increases.  

\begin{remark}
In the static deployment case \cite{yang2023analysis}, the transmission success probability of each source varies from node to node and cannot be characterized by Theorem 1.
Consequently, it requires deriving the SINR meta-distribution to characterize the link performance. 
Although this approach provides a fine-grained perspective for network analysis, it often leads to complex analytical expressions that hinder the exploration of AoI limits. 
In contrast, the mobile network model considered by this paper mitigates the influence of spatial correlations among the nodes, enabling tractable results that provide useful network design guidelines, which will be detailed in the subsequent sections. 
\end{remark}

The solution(s) to fixed-point equation \eqref{eq:define:p} can be obtained via fixed-point iterations. 
Specifically, we denote by $p_{\mathrm{s},1} \in (0, 1]$ and $p_{\mathrm{s},n}$ the initial point and estimation of transmission success probability after the $n$th iteration, respectively.
The iteration process can be expressed as follows
\begin{equation}\label{eq:define:p:it}
			p_{\mathrm{s},n+1} = \exp\left(-\frac{ \lambda c \eta r^2 }{1+A\eta p_{\mathrm{s},n}} -\frac{\theta r^{\alpha}}{\rho}\right).
\end{equation}

While the convergence of the fixed-point iteration is immune to the initial value, depending on the network configuration, different starting points may lead to different solutions. 
Therefore, we investigate the conditions of different distributions of roots of \eqref{eq:define:p} in the following.
	\begin{corollary} \label{Th:root:analysis}
		The fixed-point equation in \eqref{eq:define:p} has three distinct roots, $0<p_{\mathrm{s}}^{L}<p_{\mathrm{s}}^{M}<p_{\mathrm{s}}^{H}<1$, if  
		\begin{equation} \label{region:lcr2}
			\lambda cr^2 > \frac{4}{\eta}
		\end{equation} 
		and 
		\begin{equation} 
			A_l<A<A_h,
		\end{equation}   
  where $A_l$ and $A_h$ are given respectively as follows
		\begin{align}
			& A_l = \frac{\lambda cr^2\left(\frac{1}{2}+\sqrt{\frac{1}{4}-\frac{1}{\lambda cr^2\eta}}\right)-\frac{1}{\eta} }{\exp\left(- \frac{ \theta r^\alpha }{ \rho } -\frac{1}{\frac{1}{2}+\sqrt{\frac{1}{4}-\frac{1}{\lambda cr^2\eta}}} \right)}, \label{eq:al}\\
		& A_h = \frac{\lambda cr^2\left(\frac{1}{2}-\sqrt{\frac{1}{4}-\frac{1}{\lambda cr^2\eta}}\right)-\frac{1}{\eta} }{\exp\left(- \frac{ \theta r^\alpha }{ \rho } -\frac{1}{\frac{1}{2}-\sqrt{\frac{1}{4}-\frac{1}{\lambda cr^2\eta}}} \right)}. \label{eq:ah}
	\end{align}
    When $A=A_l$ or $A=A_h$, \eqref{eq:define:p} has two different solutions; otherwise, \eqref{eq:define:p} has a unique solution.
\end{corollary}
\begin{IEEEproof}
	Please see Appendix \ref{proofrootanalysis}. 
\end{IEEEproof}

Notably, in the scenario that \eqref{eq:define:p} has three distinct
roots, 
if $p_{ \mathrm{s}, 0 } \in (0, p_{\mathrm{s}}^{M})$ or $p_{ \mathrm{s}, 0 } \in (p_{\mathrm{s}}^{M}, 1]$, the fixed-point iteration \eqref{eq:define:p:it} results in $p_{\mathrm{s}}^{L}$ and $p_{\mathrm{s}}^{H}$, respectively.
However, regardless of the initial value of $p_{ \mathrm{s}, 0 }$, there is no guarantee from \eqref{eq:define:p:it} that the ultimate solution would be $p_{\mathrm{s}}^{M}$.
\begin{figure}[t]
	\centering
\includegraphics[width=8cm,height=6cm]{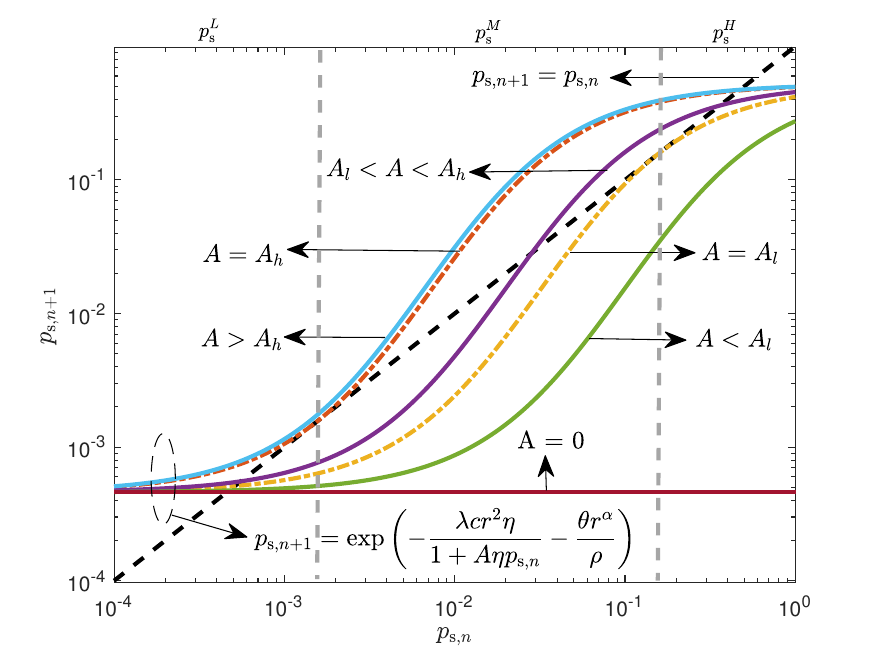}
	\caption{Approximate trajectory of $p_{\mathrm{s},n}$. Spatial deployment density $\lambda =0.15$, update rate $\eta=1$, decoding threshold $\theta=0~\text{dB}$, SNR $\rho=20~\text{dB}$, path-loss exponent $\alpha= 3.8$, and transceiver distance $r = 3$.}
	\label{fig:rootAnalysis}
 \vspace{-0.5cm}
\end{figure}

A pictorial example is provided in Fig.~\ref{fig:rootAnalysis}, where we set the parameters that satisfy $\lambda cr^2>\frac{4}{\eta}$ and showcase six distinct curves, arranged based on the value of the age threshold. Specifically, they are: $(i)$ $A=0$, $(ii)$ $A=10\in (0,A_l)$, $(iii)$ $A=A_l\approx 31$, $(iv)$ $A=50\in (A_l,A_h)$, $(v)$ $A=A_h\approx 135$, and $(vi)$ $A=150 \in (A_h, +\infty)$. 
Let us take the case of $A=50\in (A_l, A_h)$ as an example. 
In this case, the region of $p_{\mathrm{s},n}$ has four possible scenarios:  (1) $0 < p_{\mathrm{s},n}<p_{\mathrm{s}}^{L}$; (2) $p_{\mathrm{s}}^{L}<p_{\mathrm{s},n}<p_{\mathrm{s}}^{M}$; (3) $p_{\mathrm{s}}^{M}<p_{\mathrm{s},n}<p_{\mathrm{s}}^{H}$; and (4) $p_{\mathrm{s}}^{H}<p_{\mathrm{s},n}<1$.
In the first and second scenarios, $p_{\mathrm{s},n}$ will gradually approach $p_{\mathrm{s}}^{L}$. 
In contrast, $p_{\mathrm{s},n}$ tends toward $p_{\mathrm{s}}^{H}$ in the third and fourth cases.
As such, in this example, there are two candidate points, namely, $p_{\mathrm{s}}^{L}$ and $p_{\mathrm{s}}^{H}$, that $p_{\mathrm{s},n}$ can converge to.
In a similar vein, we can show that $p_{\mathrm{s},n}$ only converges to $p_{\mathrm{s}}^{H}$ (or $p_{\mathrm{s}}^{L}$) in other cases. 

In the terminology of random network analysis \cite{DaiLinALOHACSMA}, $p_{\mathrm{s}}^{L}$ and $p_{\mathrm{s}}^{H}$ are steady-state points of the fixed-point equation \eqref{eq:define:p} since they are attainable by controlling the input to the fixed-point iteration in \eqref{eq:define:p:it}. In contrast, $p_{\mathrm{s}}^{M}$ is not a steady-state point in this context.

To this end, we define the following region of network parameter configurations under which the fixed-point equation of transmission success probability has steady-state solutions, and we term these ranges the stable regions.
\begin{itemize}
	\item  \textit{Bistable region:\footnote{In the context of ALOHA-like random access networks, the bi-stable characteristic refers to a phenomenon where the system's performance (measured by, e.g., transmission success probability) can exist in one of two statistically stable equilibrium points under certain conditions\cite{BistableALOHA,DaiLinALOHACSMA,ZhanM2M, DAILINStableALOHA}. The bi-stable region is the unstable region in practice, since the system will easily converge to the state where every transmitter has a small transmission success probability. In this case, data transmissions often fail, significantly deteriorating the performance of real-time applications.}}  
   $$\mathcal{B} = \left\{\left(\lambda,\theta,r,\alpha,\rho, A, \eta\right)\Big{|}\lambda cr^2>\frac{4}{\eta}, A_l<A<A_h \right\},   
   $$
   in which the network has two steady-state points $p_{\mathrm{s}}^{L}$ and $p_{\mathrm{s}}^{H}$. 
	\item  \textit{Monostable region}:  $\mathcal{M}=\mathcal{M}_{p_{\mathrm{s}}^{H}}\cup\mathcal{M}_{p_{\mathrm{s}}^{L}}=\bar{\mathcal{B}}$ ($\bar{\mathcal{B}}$ is the complement of set $\mathcal{B}$); in this case, the network has one steady-state point, which can be further partitioned into the following.
     
     \textit{(i) High-efficiency region}: 
     \begin{small}
           \begin{equation}
             \mathcal{M}_{p_{\mathrm{s}}^{H}}=\left\{\left(\lambda,\theta,r,\alpha,\rho, A, \eta\right)\big{|}\lambda cr^2\leq\frac{4}{\eta}~\text{and}~\lambda cr^2>\frac{4}{\eta}, A\geq A_h \right\},  \notag
           \end{equation}   
     \end{small}
    in which the iteration converges to $p_{\mathrm{s}}^{H}$.
    
    \textit{(ii) Low-efficiency region:}
           \begin{equation}
             \mathcal{M}_{p_{\mathrm{s}}^{L}}=\left\{\left(\lambda,\theta,r,\alpha,\rho, A, \eta\right)\big{|}\lambda cr^2>\frac{4}{\eta}, A\leq A_l \right\},  \notag
           \end{equation}
           in which the iteration converges to $p_{\mathrm{s}}^{L}$.
\end{itemize}

The following corollary summarizes the properties of the steady-state points in terms of the configuration parameters.

\begin{corollary} \label{co:root:trend}
	The steady-state transmission success probabilities  $p_{\mathrm{s}}^{L}$ and  $p_{\mathrm{s}}^{H}$are increasing functions of the age threshold $A$, while they decrease in terms of the spatial deployment density $\lambda$ and update rate $\eta$.
\end{corollary}
\begin{IEEEproof}
Please see Appendix \ref{Proof:roottrend}. 
\end{IEEEproof}

\begin{corollary}
    The fixed-point iteration in \eqref{eq:define:p:it} always converges with a linear convergence rate.
\end{corollary}
\begin{IEEEproof}
    Please see Appendix \ref{ConvergenceRate}.
\end{IEEEproof}

It is intuitive that in densely deployed networks, the transmission success probability will benefit (more) from the TSA protocol due to its capability in mitigating channel contention. This can also be observed from Fig.~\ref{fig:rootAnalysis} by comparing the value of transmission success probability at $A=0$ with those at $A>0$.

\subsection{AoI Statistics}
Using the transmission success probability, we can derive closed-form expressions for the mean peak AoI and time-average AoI as follows.
\begin{Theorem} \label{lemma:AoI}
{
	Under the TSA protocol, the mean peak AoI over the typical link is 
	\begin{equation}
		\hat{\Delta}_0 = A + \frac{1}{\eta p_{\mathrm{s}}},  \label{eq:peakAoI}
	\end{equation}
	and the time-average AoI is
	\begin{equation} \label{eq:averageAoI}
		\bar{\Delta}_0 = \frac{A+1}{2}+ \frac{1}{\eta p_{\mathrm{s}}}-\frac{A+1}{2\left( 1+A\eta p_{\mathrm{s}}\right)},
	\end{equation}
	where $p_{\mathrm{s}}$ is given in \eqref{eq:define:p}.
 }
\end{Theorem}
\begin{IEEEproof}
Please see Appendix \ref{proofAoI}. 
\end{IEEEproof}

Leveraging the above theorem, we can investigate different regimes of the age threshold to demystify its influence on the AoI performance. 
Firstly, if we set the age threshold $A$ to be zero, the results in Theorem~\ref{lemma:AoI} reduce to the mean peak AoI and time-average AoI in large-scale networks under the SA protocol\cite{ThresholdInfocomWS}, i.e.,
\begin{equation}
	\hat{\Delta}^\mathrm{SA}_{0} = \bar{\Delta}^\mathrm{SA}_{0} = \frac{\exp\left( \lambda cr^2\eta+\theta r^\alpha \rho^{-1}\right)}{\eta}.
\end{equation}
Consequently, we can compare the AoI achieved under the SA and TSA protocols to identify the operating regimes where TSA outperforms SA.

We first compare the mean peak AoI attained under the SA and TSA protocols for $A>0$, which yields 
\begin{align}
	&\hat{\Delta}_{0}^{\mathrm{SA}}-\hat{\Delta}_{0}^{\mathrm{TSA}}=\frac{1}{\eta p_{\mathrm{s}}^{\mathrm{SA}}}-\left(A+\frac{1}{\eta p_{\mathrm{s}}^{\mathrm{TSA}}}\right)  \notag\\ 
	&=\frac{\left(\exp\left(\lambda cr^2\eta\right)-\exp\left(\frac{\lambda cr^2\eta}{1+A\eta p_{\mathrm{s}}^{\mathrm{TSA}}}\right)\right)}{ \eta \exp\left( \rho / \theta r^\alpha \right) }- A \notag \\
	& \overset{(a)}{=}\left(\lambda cr^2\left(\frac{A\eta p_{\mathrm{s}}^{\mathrm{TSA}}}{1+A\eta p_{\mathrm{s}}^{\mathrm{TSA}}}\right)  \exp\left(\lambda cr^2\eta \xi_{ \mathrm{p} } + \frac{ \theta r^\alpha }{ \rho }  \right)\right)-A\notag \\
	&\geq {\lambda cr^2\left(\frac{A\eta}{1+A\eta p_{\mathrm{s}}^{\mathrm{TSA}}}\right)}-A,
\end{align}
where $(a)$ follows from the mean value theorem, in which $\xi_{ \mathrm{p} } \in[\frac{1}{1+A\eta p_\mathrm{s}}, 1]$. 
As such, when $\lambda cr^2\geq\frac{1+A\eta p_{\mathrm{s}}^{\mathrm{TSA}}}{\eta}$, we have
\begin{equation}\label{ineq:peak}
\hat{\Delta}_{0}^{\mathrm{SA}}-\hat{\Delta}_{0}^{\mathrm{TSA}}=\lambda cr^2\left(\frac{A\eta}{1+A\eta p_{\mathrm{s}}^{\mathrm{TSA}}}\right)-A\geq 0.
\end{equation}
Following similar steps, we can compare the time-average AoI attained under the SA and TSA protocols as follows: 
\begin{align}
&\bar{\Delta}_{0}^{\mathrm{SA}}-\bar{\Delta}_{0}^{\mathrm{TSA}}>\frac{1}{\eta p_{\mathrm{s}}^{\mathrm{SA}}}-\left(\frac{A+1}{2}+\frac{1}{\eta p_{\mathrm{s}}^{\mathrm{TSA}}}\right)  \notag\\ 
	&=\frac{\left(\exp\left(\lambda cr^2\eta\right)-\exp\left(\frac{\lambda cr^2\eta}{1+A\eta p_{\mathrm{s}}^{\mathrm{TSA}}}\right)\right)}{ \eta \exp\left( \rho / \theta r^\alpha \right) }-\frac{A+1}{2} \notag \\
	& \overset{(a)}{=}\left(\lambda cr^2\left(\frac{A\eta p_{\mathrm{s}}^{\mathrm{TSA}}}{1+A\eta p_{\mathrm{s}}^{\mathrm{TSA}}}\right)  \exp\left(\lambda cr^2\eta \xi_{ \mathrm{a} } + \frac{ \theta r^\alpha }{ \rho }  \right)\right)-\frac{A+1}{2}\notag \\
	&\geq {\lambda cr^2\left(\frac{A\eta}{1+A\eta p_{\mathrm{s}}^{\mathrm{TSA}}}\right)}-{\vphantom{\lambda cr^2\left(\frac{A\eta}{1+A\eta p_{\mathrm{s}}^{\mathrm{TSA}}}\right)}\frac{A+1}{2}},
\end{align}
where $(a)$ follows from the mean value theorem, in which $\xi_{ \mathrm{a} } \in[\frac{1}{1+A\eta p_\mathrm{s}}, 1]$. 
When $ \lambda cr^2 \geq \frac{(1+A)(1+A\eta p_{\mathrm{s}}^{\mathrm{TSA}})}{2A\eta}$, we have
\begin{equation} \label{ineq:average}
\bar{\Delta}_{0}^{\mathrm{SA}}-\bar{\Delta}_{0}^{\mathrm{TSA}}>\lambda cr^2\left(\frac{A\eta}{1+A\eta p_{\mathrm{s}}^{\mathrm{TSA}}}\right)-\frac{A+1}{2}\geq0.
\end{equation}
The inequalities \eqref{ineq:peak} and \eqref{ineq:average} demonstrate that the TSA has the potential of outperforming SA in reducing AoI.

However, if we raise the age threshold to an excessively large value, i.e., $A\to\infty$, then we have 
\begin{align}
    &\hat{\Delta}_0 \sim A \rightarrow +\infty, \\
    &\bar{\Delta}_0 \sim \frac{A}{2} \rightarrow +\infty, 
\end{align}
implying that a large age threshold will be detrimental to the AoI. 

To this end, it is apparent that the age threshold, as well as the update rate, shall be adequately tuned to reveal the full potential of TSA. 
We will detail the approaches to obtain the optimal parameters in the following sections. 
For ease of exposition, we start our derivations without considering the influence of the bi-stable region.
In Section VI, we will demonstrate how to adjust the parameters to ensure network stability while concurrently optimizing the AoI performance in the presence of a bi-stable region. 

\section{Mean Peak AoI Optimization}
This section optimizes the mean peak AoI by adjusting the age threshold and update rate. 
The optimization problem can be formulated as\begin{equation}\label{eq:define:op:pro}
	\begin{split}
		&\underset{ \eta, A }{\min}\quad\hat{\Delta}_0 \\
		&~\mathrm{s.t.}\quad\eta \in (0,1], \\
		&\quad~~~~~A \in \mathbb{N}.
	\end{split}
\end{equation}
The age threshold is usually set as an integer, leading to a mixed integer programming problem in \eqref{eq:define:op:pro}, which is non-trivial to solve. 
As such, we first convert the problem into a continuous one by relaxing the age threshold to a continuous variable and then round the result to an integer.
More precisely, the relaxed optimization problem can be written as the following\begin{equation}\label{eq:define:op:pro:relax}
	\begin{split}
		&\underset{ \eta, A }{\min}\quad\hat{\Delta}_0 \\
		&~\mathrm{s.t.}\quad\eta \in (0,1], \\
		&\quad~~~~~A \in \mathbb{R^{+}}.
	\end{split}
\end{equation}
This optimization problem can be decomposed into two subproblems: 1) optimizing the update rate $\eta$ given an age threshold $A$, and 2) optimizing the age threshold $A$ given an update rate $\eta$. 

In the following, we will first investigate the solutions to these subproblems respectively to obtain a comprehensive understanding of the differences and similarities between the optimal parameters.
Then, we solve \eqref{eq:define:op:pro:relax} by jointly tuning the update rate and age threshold.

\subsection{Optimizing the Update Rate} 
The following theorem presents the optimal update rate for a given age threshold.
\begin{lemma}
\label{Theorem:op:peak:eta}
	Given an age threshold $A$, the optimal mean peak AoI $\hat{\Delta}_0^{\eta=\hat{\eta}^{*}_A}$ is given by 
	\begin{equation}
		\hat{\Delta}_0^{\eta=\hat{\eta}^{*}_{\vert A}} = 
		\begin{cases}
			\lambda cr^2 \exp\left(1+\theta r^\alpha\rho^{-1}\right), &\text{if}~\lambda cr^2>1+Ap_{s,*}\\
			A + \left(p_{\mathrm{s},*}\right)^{-1},  &\text{otherwise}
		\end{cases}
	\end{equation}
	which is achieved when the update rate $\eta$ is set to be 
	\begin{equation}  \label{equ:OptUpRt_PAoI}
		\hat{\eta}^{*}_{\vert A} = 
		\begin{cases}
			\dfrac{1}{\lambda cr^2-A\exp\left(-1-\theta r^\alpha\rho^{-1}\right)}, &\text{if}~\lambda cr^2>1+Ap_{s,*}\\
			1,  &\text{otherwise}
		\end{cases}
	\end{equation}
	where $p_{\mathrm{s},*}$ is the root of the following equation
	\begin{equation}\label{eq:pequal1}
		p_{\mathrm{s},*} = \exp\left(-\frac{\lambda cr^2}{1+Ap_{\mathrm{s},*}}-\frac{ \theta r^{\alpha} }{ \rho } \right).
	\end{equation}
\end{lemma}
\begin{IEEEproof}
   Please see Appendix \ref{prooflemma:oppeakq}. 
\end{IEEEproof}
Lemma~\ref{Theorem:op:peak:eta} shows that the optimal update rate shall be tuned to the maximum, i.e.,  $\hat{\eta}^{*}_{\vert A}=1$, when the $\lambda cr^2\leq 1+Ap_{\mathrm{s},*}$.
This is because when the nodes are sparsely deployed, each receiver has a high chance of successfully receiving updates. 
As such, the AoI performance can be effectively improved by increasing update rates. 
In contrast, when the nodes are deployed densely, each source shall decrease its update rate to circumvent channel contention and promote timely information delivery across the network.

\subsection{Optimizing the Age Threshold} 
The following theorem characterizes the optimal age threshold for a given update rate.
\begin{lemma}\label{Theorem:op:peak:threshold}
	Given an update rate $\eta$, the optimal mean peak AoI $\hat{\Delta}_0^{A=\hat{A}^{*}_{\eta}}$ is given by 
	\begin{equation}
		\hat{\Delta}_0^{A=\hat{A}^{*}_{\vert \eta}} = 
		\begin{cases}
			\lambda cr^2\exp\left(1+\frac{ \theta r^{\alpha} }{ \rho }\right), &\text{if}~~\lambda cr^2>\dfrac{1}{\eta}\\
			\frac{1}{\eta}\exp\left(\lambda cr^2 \eta + \frac{ \theta r^{\alpha} }{ \rho }\right),  &\text{otherwise}
		\end{cases}
	\end{equation}
	which is achieved when the age threshold $A$ is set to be 
	\begin{small}
		\begin{equation}\label{eq:optimal:A:peak}
			\hat{A}^{*}_{\vert \eta} = 
			\begin{cases}
				\left(\lambda cr^2 -\dfrac{1}{\eta}\right)\exp\left(1+\dfrac{\theta r^{\alpha} }{\rho}\right), &\text{if}~~\lambda cr^2>\dfrac{1}{\eta}\\
				0,  &\text{otherwise}.
			\end{cases}
		\end{equation}
	\end{small}
\end{lemma}
\begin{IEEEproof}
    Please see Appendix \ref{prooflemma2:peakA}. 
\end{IEEEproof}
Lemma 2 indicates that when the interference level is high, i.e., $\lambda cr^2\eta >1$, imposing an age threshold to the ALOHA-like protocol is beneficial for reducing mean peak AoI. 
Conversely, when the interference level is low, the conventional SA protocol is preferable in decreasing the mean peak AoI.
\subsection{Joint Optimization}  
By integrating results from the above, the theorem below presents a closed-form result for the optimal pair of the update rate and age threshold that minimizes the mean peak AoI.
\begin{Theorem}\label{Theorem:op:peak:joint}
	The optimal mean peak AoI is given by
	\begin{equation}
		\hat{\Delta}_0^{*} =
		\begin{cases}
			\lambda cr^2 \exp\left(1+\theta r^\alpha\rho^{-1}\right), &\text{if}~~\lambda cr^2>1\\
			\exp\left(\lambda cr^2+\theta r^\alpha\rho^{-1}\right),  &\text{otherwise}
		\end{cases}
	\end{equation}
	which is achieved when the age threshold and update rate together satisfy 
	\begin{equation}
		( \hat{A}^{*},\hat{\eta}^{*} )= 
		\begin{cases}
			\left(A_{1},\eta_{1}\right),  &\text{if}~~\lambda cr^2>1\\
			\left(0,1\right), &\text{otherwise}
		\end{cases}
	\end{equation}
	where $\left(A_{1},\eta_{1}\right)$ satisfy the following
	\begin{equation} \label{eq:op:paras:joint}
		A_1 = \left(\lambda cr^2- \frac{1}{\eta_1}\right)\exp\left(1+\frac{\theta r^{\alpha} }{ \rho } \right)
	\end{equation}
	and $\eta_1$ shall be confined in the range $\eta_1\in \left[\frac{1}{\lambda cr^2},1\right]$.
\end{Theorem}
\begin{IEEEproof}
 Please see Appendix \ref{proof:joint:op:peak}. 
\end{IEEEproof}
From \eqref{eq:op:paras:joint}, we observe that the optimal age threshold increases with the optimal update rate to mitigate the induced interference. 
Moreover, when $\lambda cr^2 \leq 1$, the optimal TSA configuration reduces to the conventional SA.
In addition to these, a (perhaps) more important message conveyed by Theorem~\ref{Theorem:op:peak:joint} is that by appropriately adjusting the update rate, SA can attain the same optimal mean peak AoI as TSA. 
In other words, \textit{TSA and SA are equivalent in minimizing the mean peak AoI}. 
\begin{remark}
    The optimal tuning of the update rate and the age threshold depends on the statistical traffic information (such as the spatial deployment density), instead of the real-time number of active nodes requesting transmission or prior experience. Such parameters are collected when the network is initially deployed.
\end{remark}


\section{Time-Average AoI Optimization}
This section presents the optimization of time-average AoI by solving the following optimization problem\begin{equation}\label{eq:define:op:pro:relax:ave}
	\begin{split}
		&\underset{ \eta, A }{\min}\quad\bar{\Delta}_0 \\
		&~\mathrm{s.t.}\quad\eta \in (0,1], \\
		&\quad~~~~~A \in \mathbb{N}.
	\end{split}
\end{equation}

Following similar approaches in the previous section, we relax the mixed integer programming problem \eqref{eq:define:op:pro:relax:ave} to a continuous one. Then, we decompose it into two subproblems and solve each one respectively. Finally, we integrate the whole procedure to establish a joint optimization. 
The concrete steps are detailed below. 
\subsection{Optimizing the Update Rate }
First, we optimize the update rate under a fixed age threshold, where the structural result is summarized as follows. 

\begin{lemma}\label{Theorem:op:ave:eta}
	Given an age threshold $A$, the optimal time-average AoI $\bar{\Delta}_{0}^{\eta=\bar{\eta}^{*}_{\vert A}}$ can be achieved by setting the update rate as
	\begin{small}
		\begin{equation} \label{equ:OptUpRt_AAoI}
			\bar{\eta}^{*}_{\vert A} = 
			\begin{cases}
				\dfrac{1}{\lambda cr^2-A\exp\left(-1-\theta r^\alpha\rho^{-1}\right)}, &\text{if}~\lambda cr^2>1+Ap_{\mathrm{s},*};\\
				1,  &\text{otherwise},
			\end{cases}
		\end{equation}
	\end{small}
	where $p_{s, *}$ is given by the solution to the following
	\begin{small}
		\begin{equation}\label{eq:pequal1}
			p_{\mathrm{s},*} = \exp\left(-\frac{\lambda cr^2}{1+Ap_{\mathrm{s},*}}-\dfrac{\theta r^\alpha }{\rho}\right).
		\end{equation} 
	\end{small}
	In this case, the time-average AoI is  
	\begin{small} 
		\begin{equation} \label{eq:op:eta:AAoI}
			\bar{\Delta}_0^{\eta=\bar{\eta}^{*}_{\vert A}}= 
			\begin{cases}
				\dfrac{2\left(\lambda cr^2\exp\left(1+\theta r^{\alpha}\rho^{-1}\right)\right)^2 +A(A+1)}{2\lambda cr^2\exp\left(1+\theta r^{\alpha}\rho^{-1}\right) }-A,\\
				~~~~~~~~~~~~~~~~~~~~~~~~~~~~~~~~~~~~~~~~\text{if}~\lambda cr^2>1+Ap_{\mathrm{s},*};\\
				\\
            \dfrac{A+1}{2}+\dfrac{A-1}{2(1+A p_{\mathrm{s},*})}+\dfrac{1}{p_{\mathrm{s},*}(1+A p_{\mathrm{s},*})},\\ ~~~~~~~~~~~~~~~~~~~~~~~~~~~~~~~~~~~~~~~~\text{otherwise}.
			\end{cases}
		\end{equation}
	\end{small}
\end{lemma}
\begin{IEEEproof}
Please see the Appendix \ref{proof:sole:op:eta}. 
\end{IEEEproof}
The condition that $\lambda cr^2\leq 1+Ap_{\mathrm{s},*}$ can be expressed as
\begin{equation}
p_{\mathrm{s},*} \geq \exp\left(-1-\theta r^{\alpha}\rho^{-1}\right).
\end{equation}
This inequality indicates that if the transmission success probability is lower than $\exp\left(-1-\theta r^{\alpha}\rho^{-1}\right)$ when $\eta=1$, the update rate should be reduced to ensure the transmission success probability stabilizes at this value to optimize the time-average AoI. 
Because the age threshold always contributes to the transmission success probability, the optimal update rate goes up with the increase of the age threshold.

A comparison between \eqref{equ:OptUpRt_PAoI} and \eqref{equ:OptUpRt_AAoI} reveals that for a given age threshold, the optimal update rates that minimize the mean peak AoI and time-average AoI admit the same structure. 
This finding seems surprising, as the two performance metrics have very different analytical expressions (cf. Theorem~1). And the reason boils down to that, in essence, the core of optimizing the update rate for mean peak AoI and time-average AoI lies at maximizing the spatial throughput $\eta p_{\mathrm{s}}$.


\subsection{Optimizing the Age Threshold }
In a similar vein, we explore the optimal structure of the age threshold under a fixed update rate. The following result provides a systematic approach to achieving this. 

\begin{lemma} \label{Theorem:op:ave:threshold}
	Given an update rate $\eta$, the optimal time-average AoI $\bar{\Delta}_0^{A=\bar{A}^{*}_{\vert \eta}}$ can be achieved by setting the age threshold as
	\begin{equation}\label{eq:op:A:AAoI}
		\bar{A}^{*}_{\vert \eta} = 
		\begin{cases}
			A_1, &\text{if}~~\lambda cr^2> \frac{1}{\eta}\mathbb{W}_0\left(\frac{\eta}{2}\exp\left(- \frac{ \theta r^{\alpha} }{ \rho } \right)\right); \\
			0,  &\text{otherwise},
		\end{cases}
	\end{equation}
	where $\mathbb{W}_0(\cdot)$ is  the principal branch of Lambert W function \cite{lambertW}, and $A_1$ is the single non-zero root of the following equation
	\begin{equation}\label{eq:op:AgeThre:ave1}
		\begin{split}
			&\left(1+A_1\eta p_{\mathrm{s}}\right) \left(\eta\exp\left(\frac{-\lambda cr^2\eta}{1+A_1\eta p_{\mathrm{s}}} - \frac{ \theta r^\alpha }{ \rho  } \right)-1\right)\\
			&\quad\quad -\lambda c r^2\eta\left(1+A_1\eta p_{\mathrm{s}}\right)^2 + \left(1+A_1\eta p_{\mathrm{s}}\right)^3 = \lambda c r^2\eta.
		\end{split}
	\end{equation}
\end{lemma}
\begin{IEEEproof}
Please see the Appendix \ref{prooftheorem5}. 
\end{IEEEproof}

\begin{remark}\label{remark:Op:A}
	The computational complexity (in terms of the number of iterations) in solving \eqref{eq:op:AgeThre:ave1} can be reduced by first computing $(1+A_{1}\eta p_{\mathrm{s}})$ and then calculate $p_{\mathrm{s}}$, instead of directly solving \eqref{eq:define:p} for the transmission success probability. 
	The optimal age threshold $A_1$ can be obtained by combining the numerical evaluations of $(1+A_{1}\eta p_{\mathrm{s}})$ and $p_{\mathrm{s}}$. 
\end{remark}

According to Lemma~\ref{Theorem:op:ave:threshold}, we note that the optimal age threshold is $\bar{A}^{*}_{\vert \eta} = 0$ when 
\begin{align} \label{eq:CndAgeThr}
	\lambda c r^2\leq\frac{1}{\eta}\mathbb{W}_0 \left(\frac{\eta}{2} \exp \left(- \frac{ \theta r^{\alpha} }{ \rho } \right) \right),
\end{align}
indicating that when network interference is mild, one shall not impose any waiting duration at the source nodes but employ an SA-like protocol for the status updating.
In contrast, TSA protocol is preferable when interference is severe, as it reduces concurrent transmissions, thus mitigating potential contentions. 
As a result, the nodes with urgent updating demands will benefit from a better transmission environment, which facilitates age performance. 

The condition in \eqref{eq:CndAgeThr} can be equivalently written as
\begin{equation}
	\eta \leq \frac{-\ln\left(2\lambda cr^2\exp\left(\theta r^\alpha\rho^{-1}\right) \right)}{\lambda c r^2}.
\end{equation}
This implies that TSA outperforms SA in terms of AoI when the update frequency is relatively high. 
Furthermore, 
when $\lambda cr^2>\tfrac{1}{2}\exp(-\theta r^\alpha\rho^{-1})$, TSA protocol consistently achieves a better time-average AoI than SA.
Notably, these observations are in line with those drawn in \cite{ThresholdISIT}, which are concluded based on a collision channel model.

While it is possible to solve equation \eqref{eq:op:AgeThre:ave1} numerically, obtaining a closed-form solution (which could provide design insights) is challenging. In that respect, we leverage (tight) upper and lower bounds of the time-average AoI to facilitate our optimization. 
Specifically, we can derive a lower bound to the average AoI as 
\begin{align} \label{eq:AveAoI_LB}
	\bar{\Delta}^{\text{LB}}_0 = \frac{A\eta p_{\mathrm{s}}+1}{2\eta p_{\mathrm{s}}}+\frac{1}{2\eta p_{\mathrm{s}}(1+A\eta p_{\mathrm{s}})},
\end{align}
and an upper bound as 
\begin{align} \label{eq:AveAoI_UB}
	\bar{\Delta}_0^{\text{UB}} = \frac{A\eta p_{\mathrm{s}}+1}{2\eta p_{\mathrm{s}}}+\frac{1}{2\eta p_{\mathrm{s}}(1+A\eta p_{\mathrm{s}})}+\frac{1}{2}.  
\end{align}

Then, we take a derivative of $\bar{\Delta}^{\text{LB}}_0$ (and/or $\bar{\Delta}^{\text{UB}}_0$) with respect to $A$ and assign it to zero, which gives 
\begin{equation} \label{eq:appop:AgeThre:ave1}
	(1 \!+\! A\eta p_{\mathrm{s}})^3-\lambda cr^2\eta(1 \!+\! A\eta p_{\mathrm{s}})^2 - (1 \!+\! A\eta p_{\mathrm{s}})-\lambda cr^2\eta = 0.
\end{equation}
By solving this equation, we obtain a closed-form expression for the suboptimal age threshold
\begin{equation} \label{eq:SubOptAgeThrld}
	\tilde{A}^{*}_{\vert \eta}=\frac{\Omega^{\frac{2}{3}}+(\lambda cr^2\eta-3)\Omega^{\frac{1}{3}}+(\lambda cr^2\eta)^2+3}{3\eta\Omega^{\frac{1}{3}} \exp\! \left(\dfrac{-3\lambda cr^2\eta\Omega^{\frac{1}{3}}}{\Omega^{\frac{2}{3}} \!+\! \lambda cr^2\eta\Omega^{\frac{1}{3}} \!+\! (\lambda cr^2\eta)^2 \!+\! 3} - \dfrac{ \theta r^\alpha }{\rho} \right)},
\end{equation} 
where
\begin{equation}\label{eq:omega}
	\Omega=(\lambda cr^2\eta)^3 + 18\lambda cr^2\eta+ 3\sqrt{3}\sqrt{(\lambda cr^2\eta)^4 + 11(\lambda cr^2\eta)^2 -1}.
\end{equation}

Correspondingly, we have a closed-form expression, given in \eqref{eq:asymptoticAAoI}, for the time-average AoI under the suboptimal threshold.
It is noteworthy that the difference between the bounds and the time-average AoI is at most $\frac{1}{2}-\frac{1}{2(1+A\eta p_{\mathrm{s}})}\in[0,\frac{1}{2})$, hence the suboptimal solution differs from the optimal one by (at most) a constant.
Indeed, by comparing \eqref{eq:averageAoI} and \eqref{eq:AveAoI_UB}, we can see that the time-average AoI approaches the upper bound when the age threshold $A$ becomes large.
Therefore, the solution given in \eqref{eq:SubOptAgeThrld} is asymptotically optimal. 
\begin{remark}
    \textit{Based on \eqref{eq:optimal:A:peak} and \eqref{eq:SubOptAgeThrld}, one can show that the optimal age threshold that minimizes the time-average AoI is larger than that for minimizing the mean peak AoI, since optimizing the former metric necessitates reducing the variance in the status update intervals. 
    This observation indicates that the TSA parameters need to be carefully configured when optimizing different AoI metrics.  }
\end{remark}


\begin{figure*}
	\begin{equation} \label{eq:asymptoticAAoI}
		\begin{aligned}
			\bar{\Delta}_0^{A=\tilde{A}^{*}_{\vert \eta}}=
			\frac{\left(\Omega^{\frac{2}{3}}+\lambda cr^2\eta\Omega^{\frac{1}{3}}+(\lambda cr^2\eta)^2+3\right)^2 \!\!+ 9\Omega^{\frac{2}{3}}\left(1-\eta\exp\left(\dfrac{-3\lambda cr^2\eta\Omega^{\frac{1}{3}}}{\Omega^{\frac{2}{3}} \!+\! \lambda cr^2\eta\Omega^{\frac{1}{3}} \!+\! (\lambda cr^2\eta)^2+3} -\dfrac{ \theta r^\alpha }{ \rho } \right)\right)}{6\eta\Omega^{\frac{1}{3}} \left(\Omega^{\frac{2}{3}} \!+\! \lambda cr^2\eta\Omega^{\frac{1}{3}} \!+\! (\lambda cr^2\eta)^2 \!+\! 3\right)   \exp\left(\dfrac{-3\lambda cr^2\eta\Omega^{\frac{1}{3}}}{\Omega^{\frac{2}{3}}+\lambda cr^2\eta\Omega^{\frac{1}{3}}+(\lambda cr^2\eta)^2+3} - \dfrac{ \theta r^\alpha }{ \rho } \right)}+\frac{1}{2}.
		\end{aligned}
	\end{equation}
	\hrule
 \vspace{-0.4cm}
\end{figure*}

\subsection{Joint Optimization}
The previous subsections have explored the optimization of the update rate (resp. age threshold) given the age threshold (resp. update rate). 
This part will investigate how to jointly adjust the update rate and age threshold to further reduce the time-average AoI.
\begin{algorithm}[t!]
	\caption{Iterative algorithm for AoI optimization} 
	\begin{algorithmic}[1]
		\State Input $\lambda$, $r$, $\theta$, $\rho$, $\alpha$, $n$, $\epsilon$
		\State Initialize $i = 1$ and choose $\bar{\eta}^{*}_{1}$  from (0,1] at random
		\Repeat
            \\ ***\textit{Optimize age threshold under a fixed update rate}***
            \If{$\lambda cr^2 \bar{\eta}^{*}_{i} > \mathbb{W}_0\left(\frac{\bar{\eta}^{*}_{i}}{2}\exp\left(-\theta r^{\alpha}\rho^{-1}\right) \right)$}
                \State Use Bisection method to search for $\psi=1+\bar{A}^{*}_{i}\bar{\eta}^{*}_{i} p_{\mathrm{s}}$ from \eqref{eq:op:AgeThre:ave1}, with error accuracy $\epsilon$ and in the interval $[1, 2e\lambda cr^2]$
            \State Compute $p_\mathrm{s}=\exp\left(-\lambda cr^2\psi^{-1}-\theta r^\alpha\rho^{-1} \right)$
            \State Compute $\bar{A}^{*}_{i}=(\psi-1)/{\bar{\eta}^{*}_{i}p_\mathrm{s}}$
            \State Compute $\bar{\Delta}_{0,i}$ based on $\bar{A}^{*}_{i}$, 
            $\bar{\eta}^{*}_{i}$, $p_\mathrm{s}$, and \eqref{eq:averageAoI}
            \Else
            \State Let $\bar{A}^{*}_{i}=0$
            \State Compute $\bar{\Delta}_{0,i} = \exp\left(\lambda cr^2\bar{\eta}^{*}_{i}+\theta r^\alpha\rho^{-1} \right)/\bar{\eta}^{*}_{i}$
            \EndIf
            \State $i \leftarrow i+1$
            \\ ***\textit{Optimize update rate under a fixed age threshold}***
            \State Compute $p_{\mathrm{s},*}$ base on $\eta = 1$, $\bar{A}^{*}_{i-1}$, and \eqref{eq:define:p}
		 \If{$\lambda cr^2 > 1+\bar{A}^{*}_{i-1}p_{\mathrm{s},*}$}
            \State Compute $\bar{\eta}^{*}_{i}$ based on $\bar{A}^{*}_{i-1}$, and \eqref{equ:OptUpRt_AAoI}
            \State Compute $\bar{\Delta}_{0,i}$ based on $\bar{A}^{*}_{i-1}$, and \eqref{eq:op:eta:AAoI}
            \Else
            \State Let $\bar{\eta}^{*}_{i}=1$
            \State Compute $\bar{\Delta}_{0,i}$ based on $\bar{A}^{*}_{i-1}$, and \eqref{eq:op:eta:AAoI}
            \EndIf
		\Until $|\bar{\Delta}_{0,i}-\bar{\Delta}_{0, i-1}|<\epsilon$
		\State Assign $\bar{A}^{*}\leftarrow \bar{A}^{*}_{i-1}$, $\bar{\eta}^{*}\leftarrow \bar{\eta}^{*}_{i}$ and $\bar{\Delta}_{0}^{*} \leftarrow \bar{\Delta}_{0,i}$  
		\State \Return $\bar{A}^{*}$, $\bar{\eta}^{*}$, $\bar{\Delta}_{0}^{*}$
	\end{algorithmic}
\end{algorithm}

Since the age threshold and update rate have a composite influence on the time-average AoI (cf. \eqref{eq:define:p} and \eqref{eq:averageAoI}), it is challenging to characterize the optimal values of these factors explicitly. 
As such, based on Lemmas~3 and 4, we propose an iterative approach, presented in Algorithm~1, to jointly optimize the age threshold and update rate. 
Specifically, let $i$ denote the iteration index, while $\bar{\eta}^{*}_{i}$ and $\bar{A}^{*}_{i}$ represent the update rate and age threshold in the $i$th iteration, respectively.
As detailed in Algorithm 1, we initialize $\bar{\eta}^{*}_{1}$ by a randomly chosen value from $(0,1]$, and then obtain $\bar{A}^{*}_{1}$ using Lemma~4.
Based on $\bar{\eta}^{*}_{1}$ and Lemma~3, we further calculate $\bar{A}^{*}_{2}$ and $\bar{\eta}^{*}_{2}$. 
Subsequently, we can compute each update of $\bar{\eta}_{i}^{*}$ and $\bar{A}^{*}_{i}$ for $i \geq 2$. 
The iterations repeat until reaching the termination condition $| \bar{\Delta}_{0,i}-\bar{\Delta}_{0,i-1}|<\epsilon$, where $\epsilon$ is an error accuracy.

\subsubsection{Convergence and Optimality Analysis of the Proposed Algorithms} 
Upon dividing the problem \eqref{eq:define:op:pro:relax:ave}, in Lemmas \ref{Theorem:op:ave:eta} and \ref{Theorem:op:ave:threshold}, we have demonstrated that each of the sub-problems, i.e., (V.A) and (V.B), possess a unique optimal value. Moreover, the objective function is unimodular in $A$ and $\eta$ (detailed in Appendix H and I).
Consequently, obtaining the optimal solution for each sub-problem in the iterations enables the alternative iteration algorithm to converge to the global optimum.

\subsubsection{Complexity Analysis of the Proposed Algorithms}
The sub-problem of fixing $\eta_i$ and computing $\bar{A}^{*}_{i}$ based on explicit expression \eqref{equ:OptUpRt_AAoI} has a complexity of $O(1)$. The sub-problem of fixing $\bar{A}^{*}_{i-1}$ and computing $\bar{\eta}^{*}_{i}$ has demonstrated to have a unique optimal solution in Lemma~\ref{Theorem:op:ave:eta}. 
By utilizing bisection search and the method outlined in Remark~\ref{remark:Op:A}, the complexity of this sub-problem is $O(\log_2\frac{2e\lambda cr^2}{\epsilon})$. The maximum iteration count is $n$. Consequently, the proposed algorithms have a computational complexity of $O(n \log_2\frac{2e\lambda cr^2}{\epsilon})$.

\color{black}
Furthermore, we can derive a structural form for the age threshold and update rate pair by leveraging the analytical expression in \eqref{eq:asymptoticAAoI}, which is asymptotically optimal.

\begin{Theorem}\label{Theorem:op:ave:joint}
	The optimal time-average AoI $\bar{\Delta}_0^{*}$ is achieved when the age threshold and the update rate are jointly set as
	\begin{equation}\left(\bar{A}^{*},\bar{\eta}^{*}\right)= 
		\begin{cases}
			(A_{*},1),  &\text{if}~\lambda cr^2> \mathbb{W}_0\left(\frac{1}{2}\exp\left(-\frac{ \theta r^\alpha }{\rho} \right)\right); \\
			\left(0,1\right), &\text{otherwise},
		\end{cases}
	\end{equation}
	where a closed-form approximation of  $A_{*}$ is given by 
	\begin{equation}
		A_{*} \approx \frac{\Omega_{*}^{\frac{2}{3}}+(\lambda cr^2-3)\Omega_{*}^{\frac{1}{3}}+(\lambda cr^2)^2+3}{3\Omega_{*}^{\frac{1}{3}} \exp\left(\dfrac{-3\lambda cr^2\Omega_{*}^{\frac{1}{3}}}{\Omega_{*}^{\frac{2}{3}}+\lambda cr^2\Omega_{*}^{\frac{1}{3}}+(\lambda cr^2)^2+3}-\dfrac{\theta r^\alpha }{\rho} \right)},
	\end{equation} 
	in which
	\begin{equation}
		\Omega_{*}=(\lambda cr^2)^3 + 18\lambda cr^2+ 3\sqrt{3}\sqrt{(\lambda cr^2)^4 + 11(\lambda cr^2)^2 -1}.
	\end{equation} 
	The corresponding time-average AoI is given by \eqref{eq:asymptoticAAoI:joint}.
	\begin{figure*}
		\begin{small}
			\begin{equation} \label{eq:asymptoticAAoI:joint}
				\tilde{\Delta}_0^{*}=	
				\begin{cases}
					\dfrac{\left(\Omega_{*}^{\frac{2}{3}}+\lambda cr^2\Omega_{*}^{\frac{1}{3}}+(\lambda cr^2)^2+3\right)^2 \!\!+ 9	\Omega_{*}^{\frac{2}{3}}\left(1-\exp\left(\dfrac{-3\lambda cr^2	\Omega_{*}^{\frac{1}{3}}}{	\Omega_{*}^{\frac{2}{3}} \!+\! \lambda cr^2\Omega_{*}^{\frac{1}{3}} \!+\! (\lambda cr^2)^2+3} -\dfrac{ \theta r^\alpha }{\rho} \right)\right)}{6\Omega_{*}^{\frac{1}{3}} \left(	\Omega_{*}^{\frac{2}{3}} \!+\! \lambda cr^2	\Omega_{*}^{\frac{1}{3}} \!+\! (\lambda cr^2)^2 \!+\! 3\right) \exp\left(\dfrac{-3\lambda cr^2	\Omega_{*}^{\frac{1}{3}}}{\Omega_{*}^{\frac{2}{3}}+\lambda cr^2\Omega_{*}^{\frac{1}{3}}+(\lambda cr^2)^2+3} - \dfrac{ \theta r^\alpha }{ \rho } \right)}+\dfrac{1}{2},  &\text{if}~\lambda cr^2> \mathbb{W}_0\left(\frac{\exp\left(-\frac{ \theta r^\alpha }{\rho} \right)}{2}\right),\\ 
					\exp(\lambda cr^2+ \theta r^\alpha \rho^{-1} ), &\text{otherwise}.
				\end{cases}
			\end{equation}
		\end{small}
		\hrule
	\end{figure*}
\end{Theorem}
\begin{IEEEproof}
Please see the Appendix \ref{proof:op:ave:joint}. 
\end{IEEEproof}

We can observe from the above analysis that \textit{at the optimal operating region of time-average AoI, the update rate of TSA always equals one while the age threshold varies in accordance with the interference level}. 

Theorem~\ref{Theorem:op:ave:joint} also allows us to investigate the scaling property of AoI under TSA. More precisely, when we densify the infrastructure (i.e., let $\lambda \rightarrow \infty$), the network becomes interference-limited (namely, the thermal noise is negligible), and the AoI scales as follows.

\begin{corollary} \label{coro:AoIscale}
The optimal time-average AoI under TSA protocol scales with the deployment density as
	\begin{equation}
		\lim_{\lambda cr^2 \to \infty}\frac{\bar{\Delta}_{0}^{*,\mathrm{TSA}}}{\lambda cr^2} =\frac{e}{2}, 
	\end{equation}
	in which the optimal age threshold ${A}^{*}$ satisfies
	\begin{equation}
		\lim_{\lambda cr^2 \to \infty}\frac{\bar{A}^{*}}{\lambda cr^2} = e.
	\end{equation}
	In contrast, the optimal time-average AoI under SA protocol scales as{\footnote{ The optimal mean peak AoI obeys the same scaling law under TSA and SA, and is identical to that of time-average AoI under the SA protocol. }}
	\begin{equation}
		\lim_{\lambda cr^2 \to \infty}\frac{\bar{\Delta}_{0}^{*,\mathrm{SA}}}{\lambda cr^2} = e.
	\end{equation}
\end{corollary} 

This corollary unveils that the optimal mean peak AoI and time-average AoI under TSA and SA both scale linearly with the interference level, but TSA can reduce the increasing rate of time-average AoI to half of that under conventional SA.
Note that the results in Corollary~\ref{coro:AoIscale} are obtained from the SINR model, and they coincide with the conclusions in \cite{ThresholdISIT} developed under the collision model. 
\begin{remark}
The TSA protocol retains the low computational complexity as well as low signaling overhead characteristics of ALOHA-like protocols. By introducing local AoI comparison (with the age threshold) during the channel contention period and utilizing the AoI information of the receiver carried in the ACK feedback, it forms a closed-loop control system that significantly enhances the network AoI performance.
\end{remark}

\section{AoI optimization with bi-stable Region}
In the previous derivations, we concentrate on optimizing the TSA protocol while putting aside the root distribution of the fixed-point equation in \eqref{eq:define:p}.
According to the discussions in Section III-B, if $\lambda cr^2\eta>4$, the solutions to \eqref{eq:define:p} fall in the bi-stable region, i.e., the network can stabilize in two distinct states, typically a high-efficiency state (i.e., $p_{\mathrm{s}}^{H}$) and a low-efficiency state (i.e., $p_{\mathrm{s}}^{L}$), depending on the initial conditions and the dynamics of the system. This section aims to address such an issue.

First of all, we can assess the difference between the two candidates steady state $p_{\mathrm{s}}^{H}$ and $p_{\mathrm{s}}^{L}$ according to the region described in Section III-B. By examining the cases $A=A_l$ and $A=A_h$, the feasible region for $p_{\mathrm{s}}^{H}$ and $p_{\mathrm{s}}^{L}$ can be determined as follows:
\begin{align}
& p_{\mathrm{s}}^{H} \in\left(\exp\left(-\tfrac{ \theta r^\alpha }{ \rho } -\tfrac{1}{\frac{1}{2}+\sqrt{\frac{1}{4}-\frac{1}{\lambda cr^2\eta}}} \right),~1\right), \\
&  p_{\mathrm{s}}^{L} \in \left(0,~\exp\left(-\tfrac{ \theta r^\alpha}{\rho} -\tfrac{1}{\frac{1}{2}-\sqrt{\frac{1}{4}-\frac{1}{\lambda cr^2\eta}}} \right)\right).
\end{align}
By taking a ratio between them, as
\begin{equation}
\begin{split}
 \frac{p_{\mathrm{s}}^{H}}{p_{\mathrm{s}}^{L}}&>\tfrac{\exp\left(- \tfrac{ \theta r^\alpha }{ \rho } -\tfrac{1}{\frac{1}{2}+\sqrt{\frac{1}{4}-\frac{1}{\lambda cr^2\eta}}} \right)}{\exp\left(-\tfrac{ \theta r^\alpha }{ \rho } -\tfrac{1}{\frac{1}{2}-\sqrt{\frac{1}{4}-\frac{1}{\lambda cr^2\eta}}} \right)}\\
 &=\exp\left(\sqrt{(\lambda cr^2\eta)^2-4\lambda cr^2\eta}\right). 
\end{split}
\end{equation}
This inequality demonstrates that in the bi-stable region, if the steady-state point shifts from $p_{\mathrm{s}}^{H}$ to $p_{\mathrm{s}}^{L}$, the performance loss increases exponentially with product of the interference level and update rate, i.e., $\lambda cr^2\eta$, if the parameters are not properly tuned. For example, at $\lambda cr^2\eta = 6$, the performance loss can be more than $30$ times.

In light of the above challenge, we propose parameter tuning strategies for both mean peak AoI and time-average AoI to enhance the robustness against the bi-stage region situation. 

\begin{figure}
    \centering
\includegraphics[width=8cm,height=6cm]{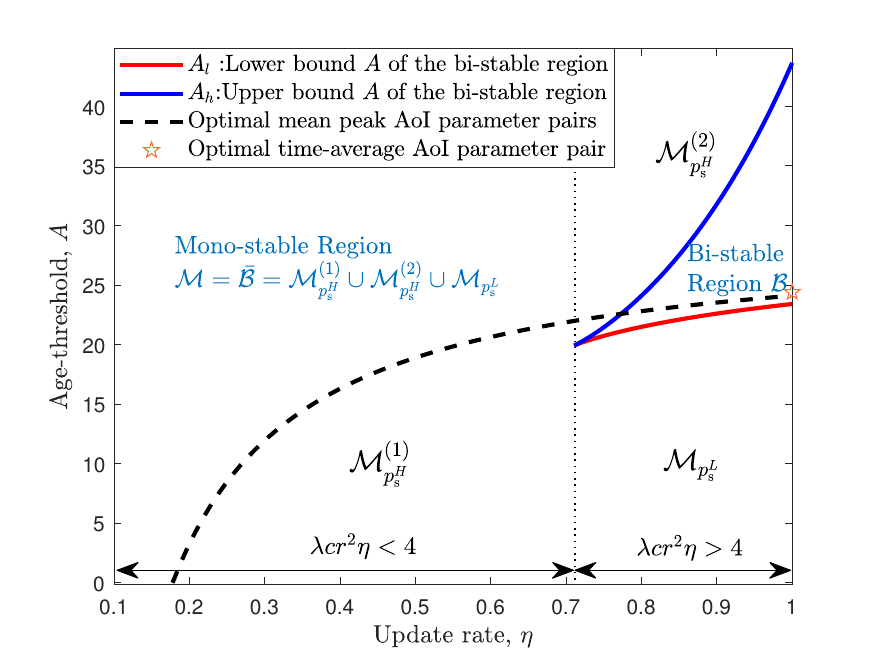}
    \caption{Bistable region $\mathcal{B}$ and monostable region $\mathcal{M}$.}
    \label{fig:Bistable}
    \vspace{-0.5cm}
\end{figure}

Fig. \ref{fig:Bistable} illustrates an example of the mono-stable and bi-stable regions to facilitate explanations of the optimization process. It also highlights instances where the optimization results from previous discussions fall within these two regions.
\subsection{Mean Peak AoI}
For the mean peak AoI optimization, the optimal parameters pairs \eqref{eq:op:paras:joint} intersect with the boundary line of the bi-stable region $A_h$. Then, the optimal parameter pairs are bisected by the curve $A_h$, with one portion lying in the mono-stable region and the other in the bi-stable region. Therefore, to achieve the optimal mean peak AoI, we only need to select optimal parameters pairs that fall within the mono-stable region. In short, the performance limits of mean peak AoI can be achieved even when the bi-stable region exists; only the tuning region for optimal parameter settings is limited. 

The region of optimal update rate $\eta_1$ when optimal mean peak AoI is achieved in \eqref{eq:op:paras:joint} can be expressed as 
\begin{equation}
    \eta_1\in \left[\frac{1}{\lambda cr^2},\min\left\{1,\frac{4.3509}{\lambda cr^2}\right\}\right].
\end{equation}
The number $4.3509$ is the approximate value of $\lambda cr^2\eta$ when curve \eqref{eq:op:paras:joint} intersects with $A_h$.
\subsection{Time-Average AoI}
For the time-average AoI optimization, the optimal performance is influenced by the bi-stable region, as the optimal operating point resides within it. Then, the  suboptimal time-average AoI is achieved when $A = A_h$,  which is given by
\begin{equation}\label{eq:constrain:op}
 \min_{\eta\in\left[\frac{4}{\lambda cr^2},1\right]}~~\frac{A_h+1}{2}+ \frac{1}{\eta p^{A=A_h}_{\mathrm{s}}}-\frac{A_h+1}{2\left(1+A_h\eta p^{A=A_h}_{\mathrm{s}}\right)}.
\end{equation}
This outcome can be intuitively interpreted as follows: 

First, due to the unstable nature of the bi-stable region, the network parameters must be fine-tuned to ensure that the system predominantly operates within the monostable region, avoiding the bi-stable region. 
Therefore, when the bi-stable region exists, the suboptimal parameters should be chosen in the boundary between the mono-stable and bi-stable regions, i.e., $A_l$ or $A_h$ (the boundary line falls within the monostable region).

Then, based on Corollary 2, we can determine that, for a given update rate, the transmission success probability on the $A_l$ curve is lower than the undesired steady-state point within the bi-stable region, and also lower than that on the $A_h$, which characterizes an extremely adverse data transmission environment. Therefore, the suboptimal parameters shall be chosen on the boundary of $A_h$ rather than $A_l$.   

Finally, upon investigating the time-average AoI performance along the boundary line $A_h$, we found that \textit{the time-average AoI decreases first and then increases as the update rate $\eta$ increases}, which means the time-average AoI has a unique optimal value in the curve of $A_h$. Then, the following gradient information will be incorporated, i.e., 
\begin{align}
    &\tfrac{\partial{\bar{\Delta}^{A=A_h} }}{\partial{\eta}}=\tfrac{1}{2}\tfrac{\partial{A_h}}{\partial{\eta}} \left(1-\tfrac{1-\eta p_\mathrm{s}^{A=A_h}}{\left(1+A_h\eta p_\mathrm{s}^{A=A_h}  \right)^2} \right) \\
    &-\left(p_\mathrm{s}^{A=A_h}+\eta \tfrac{\partial{p_\mathrm{s}^{A=A_h}}}{\partial{\eta}} \right)\left(\tfrac{1}{\left(\eta p_\mathrm{s}^{A=A_h}\right)^2} -\tfrac{A_h(A_h+1)}{2\left(1+A_h \eta p_\mathrm{s}^{A=A_h} \right)^2}\right),\notag
\end{align}
where the gradient $\frac{\partial{A_h}}{\partial{\eta}}$ and $\frac{\partial{p_\mathrm{s}^{A=A_h}}}{\partial{\eta}}$ are given as follows
\begin{align}
   &\tfrac{\partial{A_h}}{\partial{\eta}}=\tfrac{1}{\eta^2}\exp{\left(\tfrac{1}{\tfrac{1}{2}-\sqrt{\tfrac{1}{4}-\tfrac{1}{\lambda cr^2\eta}}}\right)}\exp{\left(\theta r^\alpha\rho^{-1} \right)},\\
   &\tfrac{\partial{p_\mathrm{s}^{A=A_h}}}{\partial{\eta}}=\tfrac{\lambda cr^2p_\mathrm{s}^{A=A_h}}{\lambda cr^2A_h\eta^2p_\mathrm{s}^{A=A_h}-\left(1+A_h\eta p_\mathrm{s}^{A=A_h}\right)^2}\\
   &~\times\left(1-\exp\left(\tfrac{-\lambda cr^2\eta}{1+A_h \eta p_\mathrm{s}^{A=A_h}}\right)\exp{\left(\tfrac{1}{\tfrac{1}{2}-\sqrt{\tfrac{1}{4}-\tfrac{1}{\lambda cr^2\eta}}}\right)}\right),\notag
\end{align}
the corresponding update rate in optimization problem \eqref{eq:constrain:op} can be solved effectively by numerical method.


\section{Numerical Results}
In this section, we conduct simulations to verify the accuracy of the developed analysis and then use the numerical evaluations to explore the impact of different network parameters on the AoI performance. 
Specifically, at the beginning of each simulation run, the locations of source-destination pairs are scattered over a $100\times100$ square area of unit size according to independent PPPs. 
We add the typical link to the network where the receiver is located at the center of the area. In each time slot, the locations of the links are rearranged independently according to the same PPPs except for the typical one. 
Each simulation lasts for $10^6$ time slots. We record the AoI statistics for each time slot of the typical link and then average them to obtain the mean peak and time-average AoI performance.
\begin{figure}[t]
	\centering
\includegraphics[width=8cm,height=6cm]{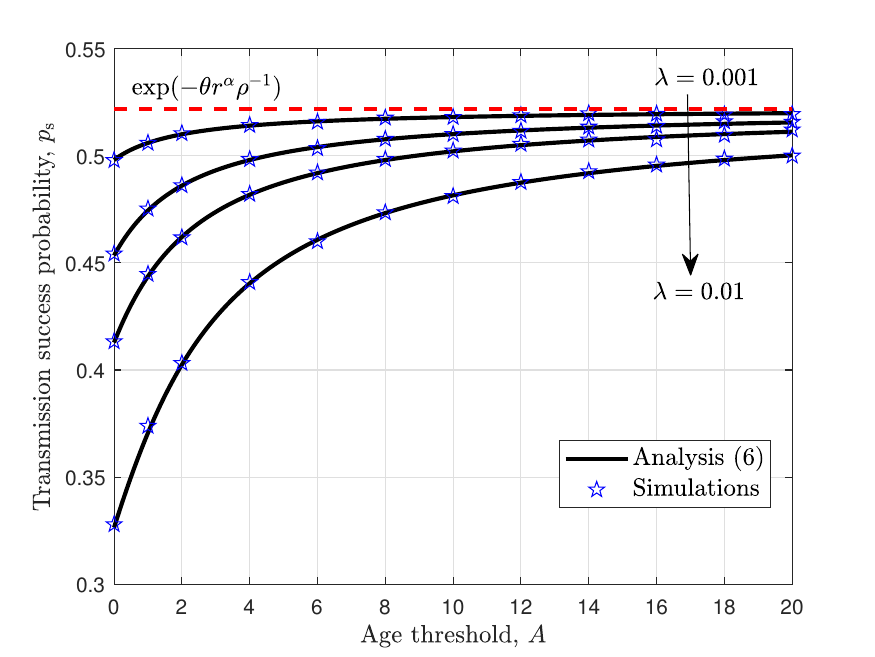}
	\caption{Transmission success probability versus age threshold. Network parameters are set as: $\eta=1$, $\lambda\in\{0.001,0.003,0.005,0.01\}$, $\theta=0~\text{dB}$, $\rho=20~\text{dB}$, $\alpha= 3.8$, and $r = 3$.}
	\label{SuccessfulTransmissionProbability}
\end{figure}

\begin{figure}[htbp]%
    \centering
    \subfigure[$A=0$ and $A=10$.]{\includegraphics[width=8cm,height=6cm]{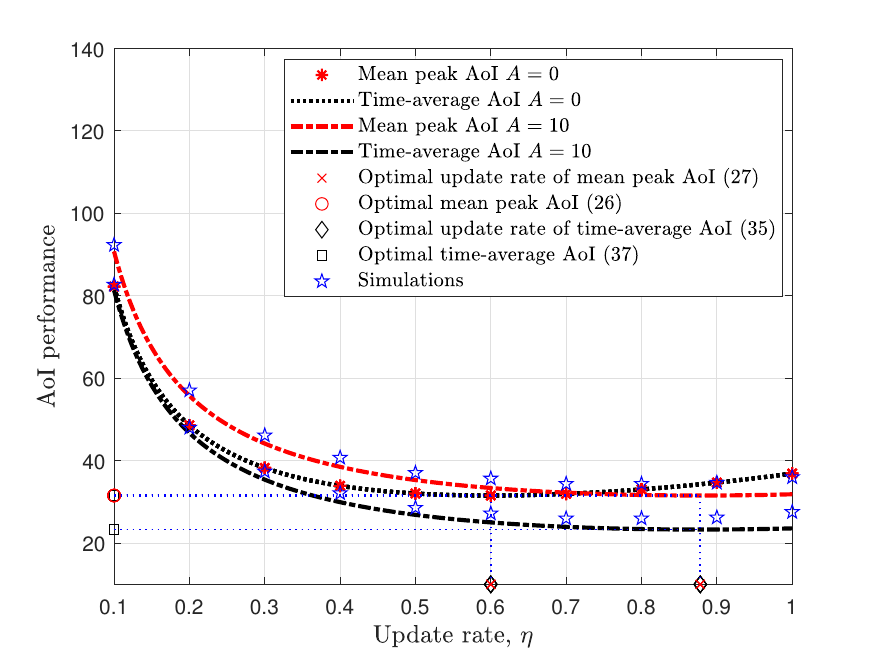}}
    \subfigure[$A=20$ and $A=50$.]{\includegraphics[width=8cm,height=6cm]{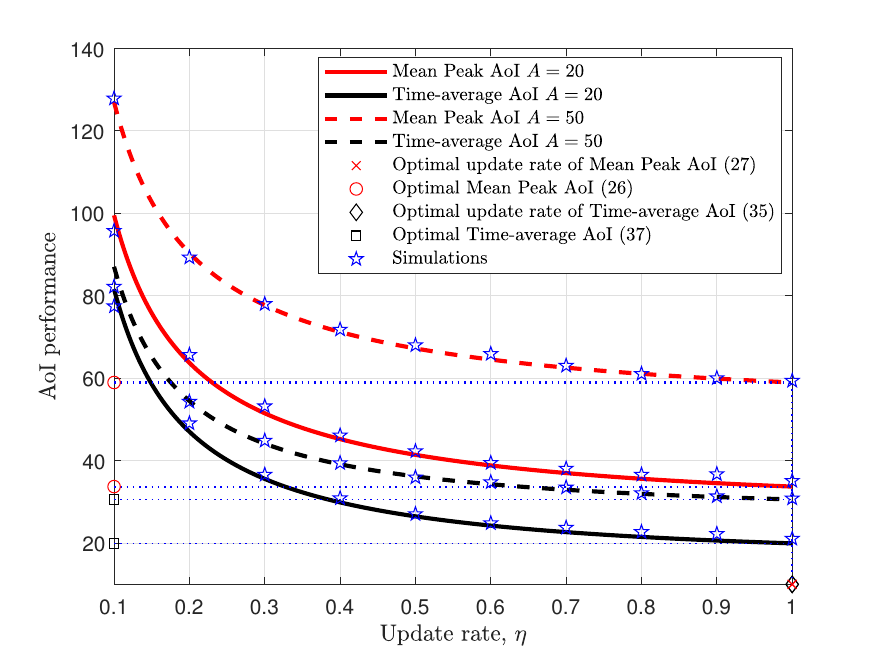}}
    \caption{AoI performance versus update rate under different age thresholds. Network
parameters are set as: $\lambda = 0.02$, $\theta=0$~dB, $\rho=20~\text{dB}$, $\alpha= 3.8$, and $r=4$.} 
  \label{AoIvsupdate}
\end{figure}

\subsection{Transmission Success Probability}
Fig.~\ref{SuccessfulTransmissionProbability} plots the transmission success probability as a function of the age threshold under a variety of spatial deployment densities. 
We notice that the analytical results and simulations are almost indistinguishable, confirming our derivations in Theorem~1. 
Moreover, under the same configuration of other network parameters, the larger the spatial deployment density,  the worse the transmission success probability due to the surging co-channel interference. 
On the other hand, the transmission success probability increases with the age threshold, demonstrating the effectiveness of the TSA protocol in interference mitigation, where the gain becomes particularly significant when the spatial deployment density is large.

\subsection{Information Freshness versus Update Rate}
Fig.~\ref{AoIvsupdate} depicts the mean peak AoI and time-average AoI as functions of the update rate under different age thresholds. 
We can see from this figure that an optimal update rate exists that minimizes the mean peak AoI and time-average AoI, whereas the optimal value of the update rate is dependent on the age threshold. 
When the age threshold is relatively large, the optimal update rate also tends to be higher. This observation corroborates the interchangeable roles of both in mitigating channel contention. 

Furthermore, we observe that the performance limits achieved by the TSA protocol in optimizing the mean peak AoI are equivalent to those achieved by the SA protocol, further substantiating this perspective. Yet, we also observe that integrating an age threshold into the SA protocol is beneficial for reducing time-average AoI, while such a threshold must be adjusted appropriately. This is attributed to the fact that, compared with adjusting the update rate, modifying the age threshold not only mitigates channel conflicts but also serves to equalize the intervals of update packet receptions. 
Consequently, it offers more pronounced gains for higher-order AoI metrics. Another noteworthy observation is that the optimal update rate (with a fixed age threshold) that realizes both the optimal mean peak AoI and the optimal time-average AoI is identical, whatever $\eta^{*}_{\vert A}<1$ in Fig.~\ref{AoIvsupdate}(a) or $\eta^{*}_{\vert A}=1$ in Fig.~\ref{AoIvsupdate}(b), corroborating the precision of our theoretical findings.
\subsection{Information Freshness versus Age Threshold}
\begin{figure}[t]
	\centering
\includegraphics[width=8cm,height=5.6cm]{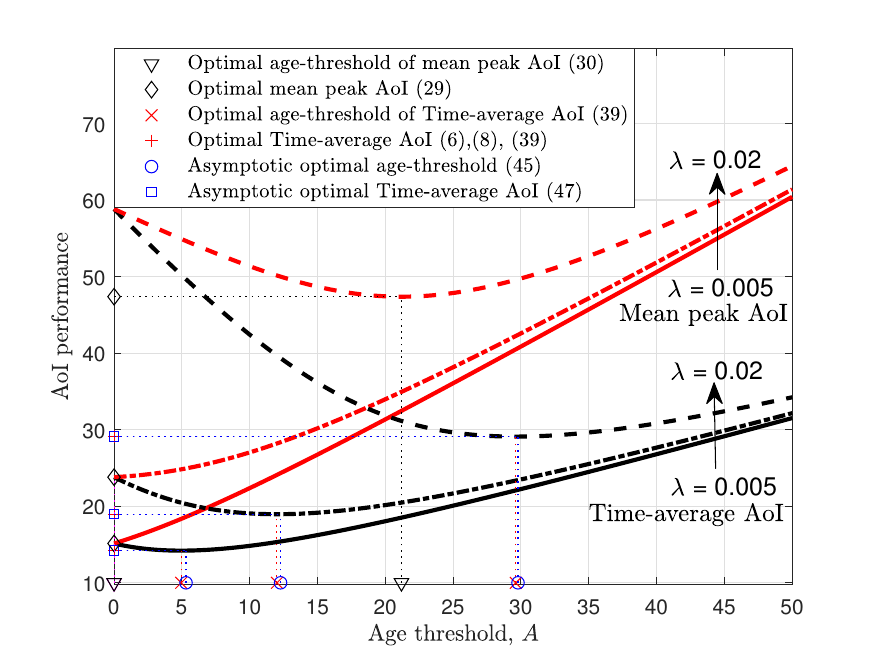}
	\caption{AoI performance versus age threshold under different deployment densities. Network
parameters are set as: $\eta=1$, $\theta=0.5$, $\rho=20~\text{dB}$, $\alpha= 3.8$, and $r=5$.}
	\label{AoIvsagethreshold}
\end{figure}

Next, Fig.~\ref{AoIvsagethreshold} plots the mean peak AoI and time-average AoI as functions of the age threshold under different spatial deployment densities. In this figure, the update rate is set to one (which is often optimal, according to Theorem~\ref{Theorem:op:ave:joint}). 
We can see that the solution given in \eqref{eq:SubOptAgeThrld} is almost indistinguishable from the optimality, confirming the accuracy of our analysis. 
Additionally, we note that the optimal age threshold increases with the spatial deployment density, owing to the more stringent requirement for interference management. Besides, the age threshold
 to achieve the optimal time-average AoI is higher
than that for the optimal mean peak AoI because of the requirement for minimal fluctuations in the data update intervals. 

\begin{figure}[t]
	\centering
	\includegraphics[width=8cm,height=5.6cm]{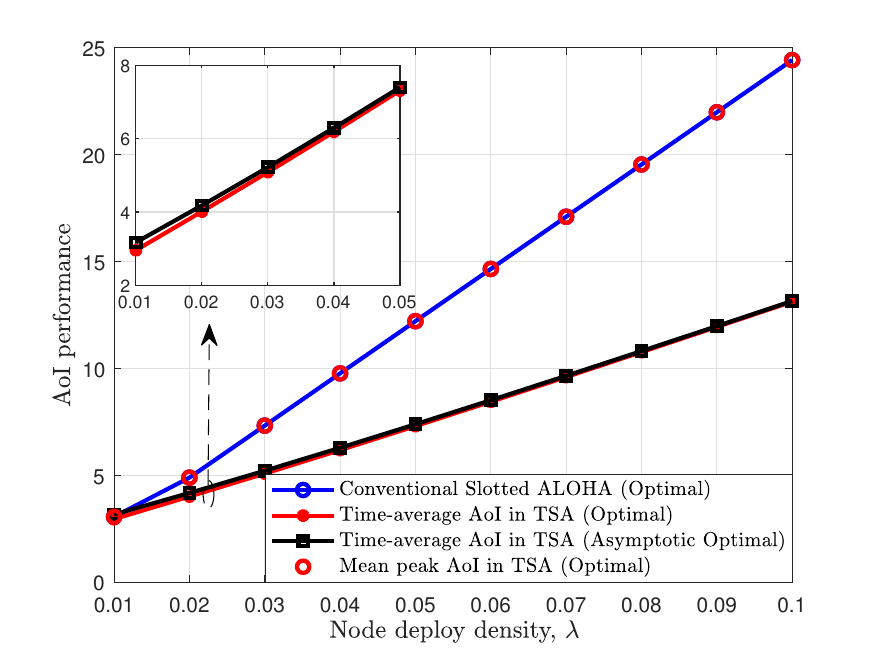}
	\caption{TSA performance gain. Network parameters are set as: $\theta= 0$~dB, $\rho=20~\text{dB}$, $\alpha= 3.8$, and $r=3$.}
	\label{fig:my_label}
\end{figure}

\begin{figure*}[t]
    \centering
    \subfigure[The first moment of the update interval]{\includegraphics[width=8cm,height=5.6cm]{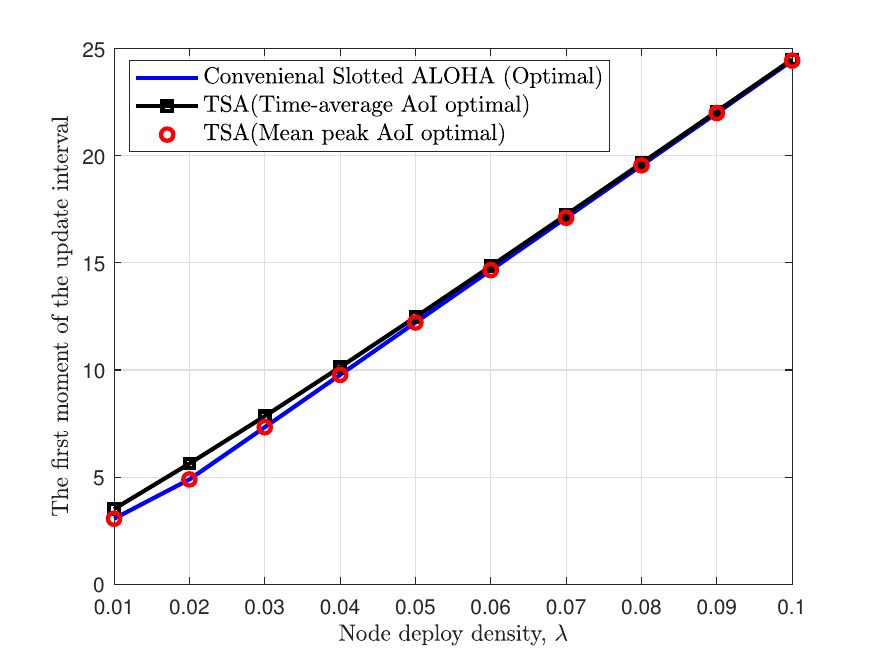}}
    \subfigure[The second moment of the update interval]{\includegraphics[width=8cm,height=5.6cm]{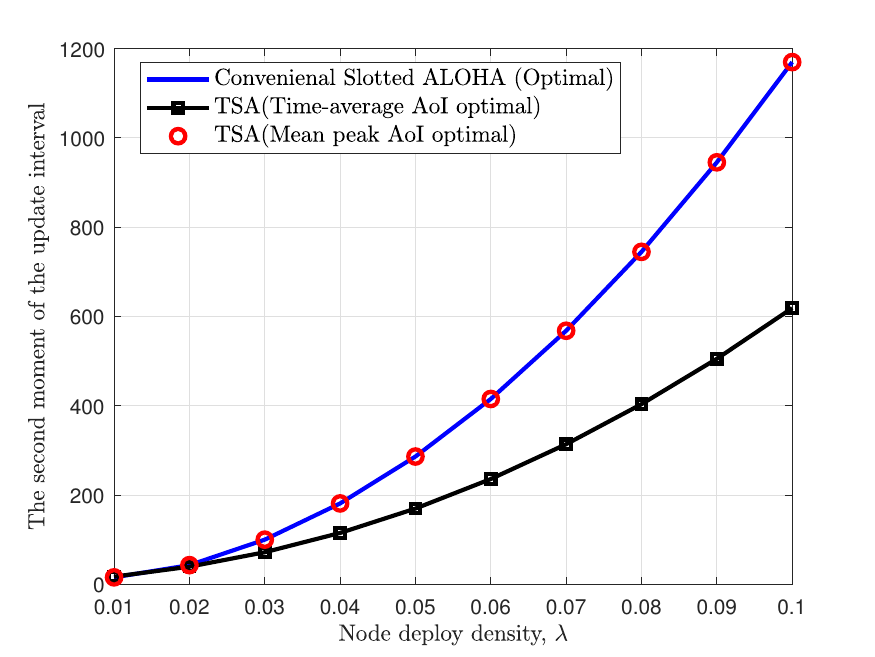}}
    \caption{TSA mechanism analysis. Network parameters are set as: $\theta= 0$~dB, $\rho=20~\text{dB}$, $\alpha= 3.8$, and $r=3$.}
    \label{fig:FunctionAnalysis}
    \vspace{-0.5cm}
\end{figure*}

\subsection{Scaling Law}
\begin{figure}[t]
    \centering
    \includegraphics[width=8cm,height=5.6cm]{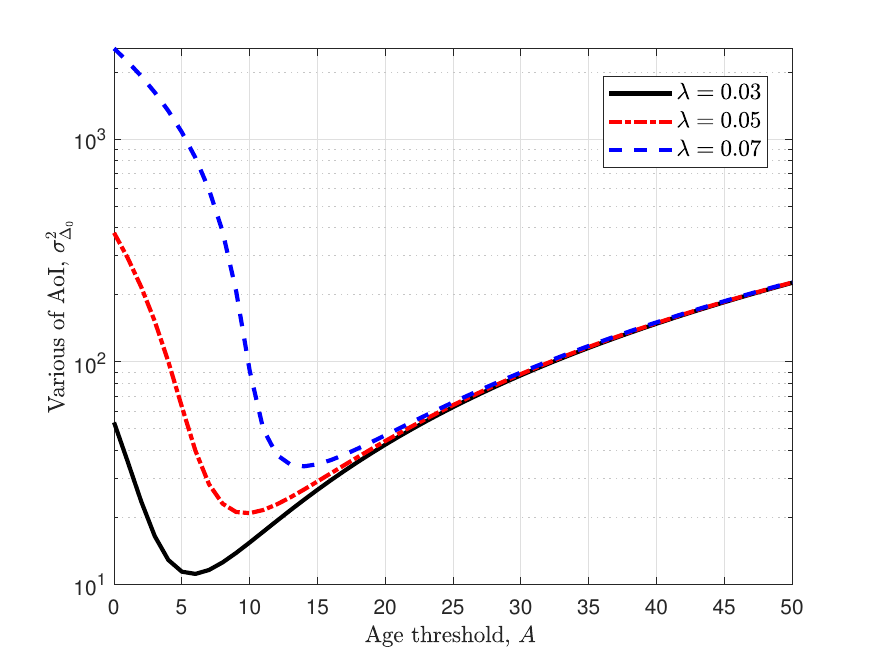}
    \caption{The variance of AoI versus age threshold under different deployment
densities. Network parameters are set as: $\theta= 0$~dB, $\rho=20~\text{dB}$, $\alpha= 3.8$, $r=3$, and $\eta=1$.}
    \label{fig:VAoI}
    \vspace{-0.5cm}
\end{figure}
Fig.~\ref{fig:my_label} illustrates the performance gain of TSA over SA by drawing the optimal time-average AoI as a function of spatial deployment density. Specifically, at each value of $\lambda$, we calculate the optimal age threshold and update rate pair using Algorithm~1 and then compute the optimal time-average AoI.
We also plot the results obtained from Theorem~\ref{Theorem:op:ave:joint} in this figure.
We can see that the optimal mean peak AoI and time-average AoI under SA and TSA increases almost linearly with the spatial deployment density, whilst the time-average AoI attained by TSA is approximately half of that under SA. 
This gain is mainly attributed to the advantage of TSA in ($a$) equalizing the status updating interval and ($b$) mitigating mutual interference, hence benefiting the AoI.
Such observation also aligns with the theoretical analysis in Corollary~\ref{coro:AoIscale}. 
We also notice that the gap between the optimal AoI and that in Theorem~\ref{Theorem:op:ave:joint} quickly diminishes, confirming the accuracy of our theoretical results.

\subsection{Higher-Order AoI Statistics}
To demystify the essential factors of TSA in reducing AoI, Fig.~\ref{fig:FunctionAnalysis} compares the first and second moments of the update interval (the time interval between two consecutive successful transmissions of the updates) under TSA and SA, where the age threshold and update rate are configured in the AoI-optimal regimes. 
Fig.~\ref{fig:FunctionAnalysis}(a) shows that when the protocol configurations are adjusted on mean peak AoI optimal, TSA and SA result in the same average length of the update interval, leading to equal mean peak AoI. 
In contrast, if the TSA configuration is set toward optimal time-average AoI, it increases the average length of the update interval (and hence the mean peak AoI), but such a difference decreases as spatial deployment density increases.  
Fig. \ref{fig:FunctionAnalysis}(b) unveils that the gain from TSA in improving time-average AoI mainly stems from controlling the fluctuation of the update interval and thus maintaining a smaller second moment of the update interval.

In addition to the average of AoI, we plot the variance of this metric \cite{FSA} in Fig.~\ref{fig:VAoI}, quantifying the deviation of the AoI performance at each link pair compared to the averaged result of the network. This figure shows that compared to SA (i.e., when $A=0$), TSA can substantially reduce the AoI variance with an adequately set age threshold (e.g., $A=10$ when $\lambda = 0.05$, which results in a reduction at the level of an order of magnitude). 
The reason can be attributed to the fact that the age threshold implicitly stipulates that the sources in good communication environments, namely, those located in sparse areas, shall remain silent for most of the time, since they have a relatively high transmission success probability; in contrast, the source nodes situated in a congested spatial area will be activated more frequently for packet transmissions, because their transmissions often fail. 
As a result, it reduces the dispersion of AoI across the network, which, in turn, enhances the fairness amongst the source nodes.

\section{Conclusion}
In this paper, we have developed an analytical framework to minimize AoI in random access networks by optimizing the TSA status updating protocol. 
We have derived analytical expressions for the transmission success probability, the mean peak AoI, and the time-average AoI of a typical link, accounting for effects from the update rate, age threshold, channel fading, and interference. Furthermore, we have obtained closed-form solutions for the optimal update rate (resp. age threshold) given a fixed age threshold (resp. update rate); we have also established a closed-form structure for the (almost) optimal update rate and age threshold pair. 
Our study showed that at the optimal operating point for time-average AoI, the update rate should be set to one while the age threshold varies according to the interference level. The optimization results have also revealed that under the same network configuration, the mean peak AoI and time-average AoI have an identical optimal update rate, yet the optimal threshold for time-average AoI is higher than that for mean peak AoI.

Furthemore, our analysis has identified the regime in which TSA outperforms SA in reducing AoI. Specifically, for the mean peak AoI, the TSA protocol exhibits comparable performance to the traditional slotted ALOHA protocol due to the interchangeable roles of tuning update rate and age threshold in mitigating channel contention. For the time-average AoI, TSA protocol outperforms SA due to its ability to equalize the intervals of update packet reception. The analysis has also disclosed the AoI scaling property, showing that the optimal mean peak AoI and time-average AoI under TSA and SA grow linearly with the deployment density; however, TSA can reduce the growth rate of time-average AoI to half of that under SA, thus, being efficient from an operation perspective.

It should be emphasized that while this study predominantly centers on the generate-at-will model, the proposed framework can also accommodate stochastic arrival patterns. Incorporating stochastic arrivals, exploring the intricate interplay among the age threshold, packet arrival, and channel access in TSA networks presents a compelling topic of discussion. Additionally, research has shown that the TSA protocol offers predictable gains for higher-order AoI metrics. However, the optimal operating point may not necessarily align perfectly with the metrics discussed in this paper. Therefore, utilizing the framework presented in this article to analyze and optimize other high-order AoI metrics with linear growth, as well as AoI metrics associated with nonlinear penalty functions, emerges as a significant direction for subsequent investigations. Moreover, enhancing the efficiency of IoT access in terms of operational costs, such as energy consumption and network lifetime, represents another promising avenue for investigation.

\appendices
\section{Proof of Theorem 1} \label{proofTheorem1}
The transmission success probability can be derived as
\begin{align}
		&\mathbb{P}(\text{SINR}_0>\theta) =\mathbb{P}\left(\frac{h_{00}r^{-\alpha}}{\sum_{i\neq 0}h_{i0}v_i ||X_i||^{-\alpha}+\rho^{-1}}>\theta \right) \notag \\
		&= \mathbb{P}\left( h_{00} > \theta r^{\alpha} \left(\sum_{i\neq 0}h_{i0}v_i ||X_i||^{-\alpha}+\rho^{-1} \right) \right) \notag \\
		& = \exp\left(-\theta r^{\alpha} \rho^{-1}\right)\prod_{i \neq 0}\exp\left( -\theta r^{\alpha} h_{i0}v_i||X_i||^{-\alpha}\right) \notag  \\
		& \overset{(a)}{=}\exp\left(-\theta r^{\alpha} \rho^{-1}\right)\prod_{i \neq 0}\left(\frac{1}{1+\theta r^{\alpha}v_i ||X_i||^{-\alpha}}\right) \notag  \\
		& \overset{(b)}{\approx} e^{-\theta r^{\alpha} \rho^{-1}} \prod_{i \neq 0}\left(1-\eta \mathbb{P}(\Delta_i(t)>A)+\frac{\eta \mathbb{P}(\Delta_i(t)>A)}{1+\theta r^{\alpha} ||X_i||^{-\alpha}}\right) \notag \\
		& = e^{-\theta r^{\alpha} \rho^{-1}} \prod_{i \neq 0}\left(1- \frac{\eta \mathbb{P}(\Delta_i(t)>A))}{1+||X_i||^{\alpha}/\left(\theta r^{\alpha}\right)}\right),
\end{align}
where (a) follows since $h_{ij}$ are i.i.d. random variables satisfy $h_{ij}\sim\exp\left(1\right)$, (b) follows as $v_i$ are independent of each other, and $\mathbb{P}(v_i=1) = \eta \mathbb{P}(\Delta_i(t)>A)$. The probability that the AoI $\Delta_i(t)$ exceed threshold $A$ can be expressed as 
\begin{equation}
	\mathbb{P}(\Delta_i(t)>A) = \frac{\frac{1}{\eta p_{\mathrm{s},i}}}{A+\frac{1}{\eta p_{\mathrm{s},i}}} = \frac{1}{1+A\eta p_{\mathrm{s},i}}, 
\end{equation}
With the mobility model, the correlation of the received SINR of each transmitter becomes i.i.d. Thus, we drop the $i$ index. The spatiotemproal interaction among the queues can be captured by a identification $p_\mathrm{s}$, which can be expressed as
	\begin{align}
		p_\mathrm{s} & =\mathbb{P}(\text{SINR}_0>\theta) \notag \\ 
		& =  e^{-\theta r^{\alpha} \rho^{-1}} \prod_{i \neq 0}\left(1- \frac{\frac{\eta}{1+A\eta p_{\mathrm{s}}}}{1+||X_i||^{\alpha}/(\theta r^{\alpha})}\right) \notag\\
		& =  e^{-\theta r^{\alpha} \rho^{-1}} \times \notag\\
		&~~~\exp\left(-\lambda \int_{x\in \mathbb{R}^2} \left[1-\left(1- \frac{\frac{\eta}{1+A\eta p_{\mathrm{s}}}}{1+||X_i||^{\alpha}/(\theta r^{\alpha})} \right) \right] dx \right) \notag \\
		& =  e^{-\theta r^{\alpha} \rho^{-1}} \exp\left(-\frac{2\pi \lambda }{\alpha}\int_{0}^{\infty}\frac{\frac{\eta}{1+A\eta p_{\mathrm{s}}}}{1+u/\theta r^{\alpha}}u^{\frac{2}{\alpha}-1}du   \right)  \notag                           \\
		& = \exp\left(-\frac{\lambda \pi\eta r^2\theta^{\frac{2}{\alpha}}\Gamma\left(1-\frac{2}{\alpha}\right)\Gamma\left(1+\frac{2}{\alpha}\right)}{1+A\eta p_\mathrm{s}} -\frac{\theta r^{\alpha}}{\rho}\right).
	\end{align}
The transmission success probability $p_\mathrm{s}$ can then be expressed as \eqref{eq:define:p} in Theorem 1.

\section{Root analysis of \eqref{eq:define:p}: Corollary 1} \label{proofrootanalysis}
We begin by constructing the following auxiliary function
\begin{equation} \label{eq:trans:fixp}
	f(p_\mathrm{s}) = -\ln p_\mathrm{s}-\frac{M}{N+p_\mathrm{s}}-K,
\end{equation}
where $M=\frac{\lambda cr^2}{A}$, $N = \frac{1}{A\eta}$, and $K=\theta r^b\rho^{-1}$.  It is evident that $f(p_\mathrm{s})=0$ is the equivalent transformation of the fixed-point equation \eqref{eq:define:p}, and can be employed to analyze the number of non-zero root for simplicity. Subsequently, we take a derivative of $f(p_\mathrm{s})$, that is  
\begin{equation} \label{eq:d1:fp}
	f^{'}(p_\mathrm{s}) = \frac{\phi(p_\mathrm{s})}{p_\mathrm{s}(N+p_\mathrm{s})^2}.
\end{equation}
The numerator of $f^{'}(p_\mathrm{s})$, i.e., $\phi(p_\mathrm{s})$ is a quadratic function, which can be expressed as 
\begin{equation}\label{eq:d1:fp:phi}
	\phi(p_\mathrm{s}) = -\left(p_\mathrm{s}+N-\frac{M}{2}\right)^2+\frac{M^2}{4}-MN.
\end{equation}
The following lemma demonstrates that the number of non-zero roots of \eqref{eq:d1:fp} is determined by the zero-points of roots of \eqref{eq:d1:fp:phi}. 
\begin{lemma} \label{lemma:fp:roots}
	$f(p_\mathrm{s})$ exhibits three non-zero roots of $0<p_{\mathrm{s}}^{L}<p_{\mathrm{s}}^{M}<p_{\mathrm{s}}^{H}<1$, if and only if $\phi(p_\mathrm{s})=0$ has two non-zero roots $0<p^{'}_{\mathrm{s},1}<p^{'}_{\mathrm{s},2}<1$ with $f(p^{'}_{\mathrm{s},1}) <0$ and $f(p^{'}_{\mathrm{s},2}) >0$; in the cases where $f(p^{'}_{\mathrm{s},1})=0$ or $f(p^{'}_{\mathrm{s},2})=0$; $f(p_\mathrm{s})=0$ has two non-zero roots, otherwise, $f(p_\mathrm{s})=0$ will solely have one non-zero root $0<p_\mathrm{s}<1$.
\end{lemma}

\begin{IEEEproof}
Due to the fact $\underset{p_\mathrm{s}\to 0}{\lim}f(p_\mathrm{s})>0$, $f(1)=-\frac{M}{N+1}-K<0$, $f(p_\mathrm{s})$ is a continuous function, according to the zero-point theorem, $f(p_\mathrm{s})$ possesses a non-zero root at least. Furthermore, the denominator of the function $f^{'}(p_\mathrm{s})$ is greater than zero for $p_\mathrm{s}\in(0,1]$. Consequently, the number of non-zero roots of $\phi(p_\mathrm{s})=0$ for
	$p_\mathrm{s}\in(0,1]$ determines the number of non-zero roots of $f^{'}(p_\mathrm{s})=0$ for $p_\mathrm{s}\in(0,1]$, and we will only consider the root of $\phi(p_\mathrm{s})$ in the following scenarios.
	
	(1) Assume that $\phi(p_\mathrm{s})=0$ has no non-zero root when $p_\mathrm{s}\in(0,1]$, it follows that in this case $\phi(p_\mathrm{s})<0$ for $p_\mathrm{s}\in (0,1]$. Consequently, $f^{'}(p_\mathrm{s})<0$ and $f(p_\mathrm{s})$ decreases monotonically when $p_\mathrm{s}\in(0,1]$, Hence, $f(p_\mathrm{s})$ has only one non-zero root in this scenario.
	
	(2) Assume that $\phi(p_\mathrm{s})=0$ has one non-zero root $0<p_\mathrm{s}^{'}\leq1$, we have the following scenarios:
	
	~~(2.a) If $\phi(p_\mathrm{s})=0$ has two roots when $p_\mathrm{s}\in \mathbb{R}$, and one of them in range $p_\mathrm{s}\in(0,1)$, we denote the root in $(0,1)$ as $p_\mathrm{s}^{'}$. Consequently, when $p_\mathrm{s}\in (0,p_\mathrm{s}^{'})$, $f(p_\mathrm{s})$ monotonically decrease and $p_\mathrm{s}\in (p_\mathrm{s}^{'},1)$, $f(p_\mathrm{s})$ monotonically increase. Since we have $f(1)<0$, thus $f(p_\mathrm{s})$ only has one non-zero root.
	
	~~(2.b) If $\phi(p_\mathrm{s})=0$ has one root when $p_\mathrm{s}\in \mathbb{R}$, and this root in range $p_\mathrm{s}\in(0,1)$, then $f(p_\mathrm{s})$ monotonically decrease when $p_\mathrm{s}\in(0,1)$, so $f(p_\mathrm{s})$ only has one non-zero root. Combining two cases, when $\phi(p_\mathrm{s})=0$ has one non-zero root, $f(p_\mathrm{s})=0$ only has one non-zero root.

     Combining these two cases, when $\phi(p_\mathrm{s})=0$ has one non-zero root, $f(p_\mathrm{s})=0$ only has one non-zero root.
     
	(3) Assume that $\phi(p_\mathrm{s})=0$ has two non-zero roots $0<p^{'}_{\mathrm{s},1}<p^{'}_{\mathrm{s},2}<1$, then we have
	
	~~(3.a) If $p_{\mathrm{s},2}^{'}=1$, then $\phi(p_\mathrm{s})<0$ for $p_\mathrm{s}\in(0,p_{\mathrm{s},1}^{'})$ and $\phi(p_\mathrm{s})>0$ for $p_\mathrm{s}\in(p_{\mathrm{s},1}^{'},1)$. Consequently, $f^{'}(p_\mathrm{s})<0$ for $p_\mathrm{s}\in(0,p_{\mathrm{s},1}^{'})$ and $f^{'}(p_\mathrm{s})>0$ for $p_\mathrm{s}\in(p_{\mathrm{s},1}^{'},1)$, indicating that $f(p_\mathrm{s})$ monotonically decreases for $p_\mathrm{s}\in(0,p_{\mathrm{s},1}^{'})$, and increases for $p_\mathrm{s}\in(p_{\mathrm{s},1}^{'},1]$. Since $f(1)<0$, we can conclude that in this case, $f(p_\mathrm{s})=0$ only one non-zero root $0<p_{\mathrm{s}}^{H}<1$.
	
    ~~(3.b) If $p^{'}_{\mathrm{s},2}<1$, then $\phi(p_\mathrm{s})<0$ for $p_\mathrm{s}\in(0, p^{'}_{\mathrm{s},1})\cup(p^{'}_{\mathrm{s},2}, 1)$, and $\phi(p_\mathrm{s})>0$ for $p_\mathrm{s}\in (p^{'}_{\mathrm{s},1}, p^{'}_{\mathrm{s},2})$. As a result, $f^{'}(p_\mathrm{s})<0$ for $p_\mathrm{s}\in(0, p^{'}_{\mathrm{s},1})\cup(p^{'}_{\mathrm{s},2}, 1)$, and $f^{'}(p_\mathrm{s})>0$ for $p_\mathrm{s}\in (p^{'}_{\mathrm{s},1}, p^{'}_{\mathrm{s},2})$, indicating that $f(p_\mathrm{s})$ monotonically decreases for $p_\mathrm{s}\in(0, p^{'}_{\mathrm{s},1})\cup(p^{'}_{\mathrm{s},2}, 1)$, and increases for $p\in (p^{'}_1, p^{'}_2)$.Then, we have if $f(p_{\mathrm{s},1}^{'})>0$ or $f(p_{\mathrm{s},2}^{'})<0$, $f(p_\mathrm{s})=0$ has one root
	$0<p_{\mathrm{s}}^{H}\leq1$; when $f(p^{'}_{\mathrm{s},1})=0$ or $f(p^{'}_{\mathrm{s},2})=0$; $f(p_\mathrm{s})=0$ has two non-zero roots; otherwise, $f(p_\mathrm{s})=0$ has three non-zero roots $0<p_{\mathrm{s}}^{L}<p_{\mathrm{s}}^{M}<p_{\mathrm{s}}^{H}\leq 1$ in which $f(p_{\mathrm{s},1}^{'})<0$ and $f(p_{\mathrm{s},2}^{'})>0$.
\end{IEEEproof}

\color{black}
Lemma 4 further concludes the specific condition for
$\phi(p_\mathrm{s}) = 0$ has two non-zero roots  $0<p^{'}_{\mathrm{s},1}<p^{'}_{\mathrm{s},2}<1$ with $f(p^{'}_{\mathrm{s},1}) <0$ and $f(p^{'}_{\mathrm{s},2}) >0$.
\begin{lemma} \label{lemma:threeroots:cond}
	$\phi(p_{\mathrm{s}})=0$ has two non-zero roots $0<p^{'}_{\mathrm{s},1}<p^{'}_{\mathrm{s},2}<1$ with $f(p^{'}_{\mathrm{s},1}) <0$ and $f(p^{'}_{\mathrm{s},2}) >0$ if and only if  $\lambda cr^2>\frac{4}{\eta}$, and $A_l<A<A_h$, where $A_l$ and $A_h$ are respectively given by
  \begin{align}
			& A_l = \frac{\lambda cr^2\left(\frac{1}{2}+\sqrt{\frac{1}{4}-\frac{1}{\lambda cr^2\eta}}\right)-\frac{1}{\eta} }{\exp\left(- \frac{ \theta r^\alpha }{ \rho } -\frac{1}{\frac{1}{2}+\sqrt{\frac{1}{4}-\frac{1}{\lambda cr^2\eta}}} \right)}; \label{eq:al}\\
		& A_h = \frac{\lambda cr^2\left(\frac{1}{2}-\sqrt{\frac{1}{4}-\frac{1}{\lambda cr^2\eta}}\right)-\frac{1}{\eta} }{\exp\left(- \frac{ \theta r^\alpha }{ \rho } -\frac{1}{\frac{1}{2}-\sqrt{\frac{1}{4}-\frac{1}{\lambda cr^2\eta}}} \right)}.\label{eq:ah}
	\end{align}
\end{lemma}
\begin{IEEEproof}
	Considering the properties of quadratic function, $\phi(p_\mathrm{s})=0$ has two non-zero roots when $\underset{p_\mathrm{s}\to 0}{\lim}~\phi(p_\mathrm{s})<0$, $\underset{p_\mathrm{s}\to 1}{\lim}~\phi(p_\mathrm{s})<0$ and the peak value of $\phi(p_\mathrm{s})$ is larger than zero, that is: 
	\begin{align}
		&~\underset{p_\mathrm{s}\to 0}{\lim}~\phi(p_\mathrm{s})= -N^2<0  \label{eq:p0}\\
		&~\underset{p_\mathrm{s}\to 1}{\lim}~\phi(p_\mathrm{s})= -\left(1+N-\frac{M}{2}\right)^2+\frac{M^2}{4}-MN<0 \label{eq:p1}\\
		&\underset{p_\mathrm{s}\in(0,1)}{\text{max}}~\phi(p_\mathrm{s}) = \frac{M^2}{4}-MN>0 \label{eq:ppeak}
	\end{align}
	Combining \eqref{eq:p0}, \eqref{eq:p1} and \eqref{eq:ppeak}, we have
	\begin{equation} \label{eq:reg:lcr}
		\frac{4}{\eta}<\lambda cr^2 < \min \left\{A\left(\frac{1}{A\eta}+1\right)^2, 2A+\frac{2}{\eta} \right\}
	\end{equation}
	For $\frac{4}{\eta} < \frac{2}{\eta}+2A$, we have $\frac{1}{A\eta^2}-A<0$ . Then, 
	\begin{equation}
		A\left(\frac{1}{A\eta}+1\right)^2 - \left(2A+\frac{2}{\eta}\right) = \frac{1}{A\eta^2}-A<0
	\end{equation}
	which means \eqref{eq:reg:lcr} can be simplified as
	\begin{equation} \label{eq:lcr2:bound} 
		\frac{4}{\eta}<\lambda cr^2 < A\left(\frac{1}{A\eta}+1\right)^2
	\end{equation}
	Then, we consider the condition that $f(p^{'}_{\mathrm{s},1}) <0$ and $f(p^{'}_{\mathrm{s},2}) >0$, i.e.,
 \begin{small}
      \begin{align}\label{eq:fp1'MN} 
  f(p^{'}_{\mathrm{s},1})&=-\frac{M}{\frac{{M}}{2}-\sqrt{\frac{{M}^2}{4}-MN}}-\ln\left(\frac{M}{2}-N-\sqrt{\frac{{M}^2}{4}-MN}\right) \notag\\
  &-\theta r^\alpha\rho^{-1} < 0,
\end{align}%
 \end{small}
		and
         \begin{small}
		\begin{align}	\label{eq:fp2'MN}	f(p^{'}_{\mathrm{s},2})&=-\frac{M}{\frac{M}{2}+\sqrt{\frac{{M}^2}{4}-MN}}-\ln\left(\frac{M}{2}-N+\sqrt{\frac{{M}^2}{4}-MN}\right)\notag\\&-\theta r^\alpha\rho^{-1} > 0,
		\end{align}
        \end{small}
	By solving the inequalities \eqref{eq:fp1'MN} and \eqref{eq:fp2'MN}, we then have 
    \begin{equation}
        A>A_l = \frac{\lambda cr^2\left(\frac{1}{2}+\sqrt{\frac{1}{4}-\frac{1}{\lambda cr^2\eta}}\right)-\frac{1}{\eta} }{\exp\left(- \frac{ \theta r^\alpha }{ \rho } -\frac{1}{\frac{1}{2}+\sqrt{\frac{1}{4}-\frac{1}{\lambda cr^2\eta}}} \right)}
    \end{equation}
    and 
    \begin{equation}
      A<A_h = \frac{\lambda cr^2\left(\frac{1}{2}-\sqrt{\frac{1}{4}-\frac{1}{\lambda cr^2\eta}}\right)-\frac{1}{\eta} }{\exp\left(- \frac{ \theta r^\alpha }{ \rho } -\frac{1}{\frac{1}{2}-\sqrt{\frac{1}{4}-\frac{1}{\lambda cr^2\eta}}} \right)}. 
    \end{equation}
	Then, by solving the inequality $\lambda c R^2<A(\frac{1}{A\eta}+1)^2$,
	we have 
	\begin{equation}\label{eq:Abound1}
		A>\lambda cr^2\left(\frac{1}{2}+\sqrt{\frac{1}{4}-\frac{1}{\lambda cr^2\eta}}\right)-\frac{1}{\eta},
	\end{equation}
	or 
	\begin{equation}\label{eq:Abound2}
		A<\lambda cr^2\left(\frac{1}{2}-\sqrt{\frac{1}{4}-\frac{1}{\lambda cr^2\eta}}\right)-\frac{1}{\eta}
	\end{equation}
	When the condition $A_l<A<A_h$ is satisfied, the inequality \eqref{eq:Abound1} must be satisfied, and the inequality \eqref{eq:Abound2} will not be satisfied. Then, the condition \eqref{eq:lcr2:bound} can be simplified as $\lambda cr^2>\frac{4}{\eta}$.
\end{IEEEproof}
Consequently, Corollary 1 can be proved by combining Lemma \ref{lemma:fp:roots} and lemma \ref{lemma:threeroots:cond}.

\section{Proof of Monotony of $p^{L}_{\mathrm{s}}$ and $p^{H}_{\mathrm{s}}$: Corollary 2} \label{Proof:roottrend}
By taking a partial derivative of $p_\mathrm{s}$ with respect to $A$, $\eta$, and $\lambda$, we have
	\begin{align}
		&	\frac{\partial{p_\mathrm{s}}}{\partial{A}} = \frac{\lambda cr^2\eta^2p_\mathrm{s}^2}{(1+A\eta p_\mathrm{s})^2-\lambda cr^2A\eta^2p_\mathrm{s}} =\frac{-\lambda cr^2 p_\mathrm{s}^2}{A^2} \frac{1}{\phi(p_\mathrm{s})} \label{eq:d:A:p}; \\
		&	\frac{\partial{p_\mathrm{s}}}{\partial{\eta}} = \frac{-\lambda cr^2p_\mathrm{s}}{(1+A\eta p_\mathrm{s})^2-\lambda cr^2A\eta^2p_\mathrm{s}} = \frac{\lambda cr^2 p_\mathrm{s}}{A^2 \eta^2} \frac{1}{\phi(p_\mathrm{s})}  \label{eq:d:eta:p};  \\
		&	\frac{\partial{p_\mathrm{s}}}{\partial{\lambda}} =  \frac{-cr^2\eta p_\mathrm{s}(1+A\eta p_\mathrm{s})}{(1+A\eta p_\mathrm{s})^2-\lambda cr^2A\eta^2p_\mathrm{s}} =  \frac{cr^2\eta p_\mathrm{s}(1+A\eta p_\mathrm{s})}{A^2\eta^2}  \frac{1}{\phi(p_\mathrm{s})},  \label{eq:d:lambda:p}
	\end{align}
where $\phi(p_\mathrm{s})$ is consistence with equation \eqref{eq:d1:fp:phi}. Then we prove $\phi(p_\mathrm{s}^{L})$ and $\phi(p_\mathrm{s}^{H})$ both smaller than zero. 
\begin{itemize}
	\item[(1)] If $\phi(p_\mathrm{s})=0$ has no root for $p_\mathrm{s}\in(0,1]$, then $\phi(p_\mathrm{s})<0$ for $p_\mathrm{s}\in (0,1]$. In this case, \eqref{eq:define:p} has one-zero root $p_{\mathrm{s}}^{H}$, and we have $\phi(p_{\mathrm{s}}^{H})<0$.
	
	\item[(2)] If $\phi(p_\mathrm{s})=0$ has one root for $0<p^{'}_{\mathrm{s},1}\leq1$, we have:
	\begin{itemize}
		\item[(2.a)] If $\phi(p_\mathrm{s})=0$ has two roots when $p_\mathrm{s}\in \mathbb{R}$, and one of them in range $p_\mathrm{s}\in(0,1]$, then $\phi(p_\mathrm{s})<0$ for $p_\mathrm{s}\in(0,p^{'}_{\mathrm{s},1})$ and $\phi(p_\mathrm{s})>0$ for $p_\mathrm{s}\in(p^{'}_{\mathrm{s},1},1)$. The fixed point equation \eqref{eq:define:p} has one-zero root $p_\mathrm{s}^{H}$, and $p_\mathrm{s}^{H}< p^{'}_{\mathrm{s},1}$, we then have $\phi(p_\mathrm{s}^{H})<0$.
		
		\item[(2.b)] If $\phi(p_\mathrm{s})=0$ has one root when $p_\mathrm{s} \in \mathbb{R}$, and this root in range $p_\mathrm{s}\in(0,1)$, then $\phi(p_\mathrm{s})<0$ for $p_\mathrm{s}\in (0,1]$. \eqref{eq:define:p} has one-zero root $\phi(p_\mathrm{s}^{H})$. We have $\phi(p_\mathrm{s}^{H})<0$.
		
	\end{itemize}
	\item[(3)] If $\phi(p_\mathrm{s})=0$ has two roots that $0<p^{'}_{\mathrm{s},1}<p^{'}_{\mathrm{s},2}<1$, two scenarios as follows:
	\begin{itemize}
		\item [(3.a)] If $p^{'}_{\mathrm{s},2}=1$, then $\phi(p_\mathrm{s})<0$ for $p_\mathrm{s}\in(0,p^{'}_{\mathrm{s},1})$ and $\phi(p_\mathrm{s})>0$ for $p_\mathrm{s}\in(p^{'}_{\mathrm{s},1},1)$, \eqref{eq:define:p} has one root $p_{\mathrm{s}}^{H}<p^{'}_{\mathrm{s},1}$. We then have $\phi(p^{H}_{\mathrm{s}})<0$.
		
		\item [(3.b)] If $p^{'}_{\mathrm{s},2}<1$, then $\phi(p_\mathrm{s})<0$ for $p_\mathrm{s}\in(0, p^{'}_{\mathrm{s},1})\cup(p^{'}_{\mathrm{s},2}, 1)$, and $\phi(p_\mathrm{s})>0$ for $p_\mathrm{s}\in (p^{'}_{\mathrm{s},1}, p^{'}_{\mathrm{s},2})$. In this case, \eqref{eq:define:p} may has one root $p_\mathrm{s}^{H}\in(0, p^{'}_{\mathrm{s},1})\cup(p^{'}_{\mathrm{s},2}, 1)$, or three roots $\phi(p_\mathrm{s}^{L})<p^{'}_{\mathrm{s},1}<\phi(p_\mathrm{s}^{M})<p^{'}_{\mathrm{s},2}<\phi(p_\mathrm{s}^{H})$. We then have $\phi(p_\mathrm{s}^{L})<0$ and $\phi(p_\mathrm{s}^{H})<0$.
	\end{itemize}
\end{itemize}

In summary, $\phi(p_\mathrm{s}^{L})<0$ and $\phi(p_\mathrm{s}^{H})<0$ can be proved by combining above cases. Thus, it can be obtained from \eqref{eq:d:A:p} -- \eqref{eq:d:lambda:p} that $\frac{\partial{p_\mathrm{s}}}{\partial{A}}>0$, $\frac{\partial{p_\mathrm{s}}}{\partial{\eta}}<0$ and $\frac{\partial{p_\mathrm{s}}}{\partial{\lambda}}<0$ at both the steady states $p_\mathrm{s}^{L}$ and $p_\mathrm{s}^{H}$. Therefore, the denominator of derivation satisfied the following inequality: 
\begin{equation}
	(1+A\eta p_\mathrm{s})^2-\lambda cr^2A\eta^2p_\mathrm{s}>0.
\end{equation}

\section{Convergence of \eqref{eq:define:p:it}: Corollary 3}\label{ConvergenceRate}
By examining the cases $A=A_l$ and $A=A_h$, the feasible region for $p_{\mathrm{s}}^{H}$ and $p_{\mathrm{s}}^{L}$ can be determined as follows:
\begin{align}
& p_{\mathrm{s}}^{H} \in\left(\exp\left(-\tfrac{ \theta r^\alpha }{ \rho } -\tfrac{1}{\frac{1}{2}+\sqrt{\frac{1}{4}-\frac{1}{\lambda cr^2\eta}}} \right),~1\right), \\
&  p_{\mathrm{s}}^{L} \in \left(0,~\exp\left(-\tfrac{ \theta r^\alpha}{\rho} -\tfrac{1}{\frac{1}{2}-\sqrt{\frac{1}{4}-\frac{1}{\lambda cr^2\eta}}} \right)\right).
\end{align}

Without loss of generality, we consider $p_\mathrm{s}^{H}$, which exists when (1) $\lambda cr^2\eta\leq 4$ and (2) $\lambda cr^2\eta > 4$ and $A > A_l$. 
For $A>0$, we assume the initial iterative value $p_{\mathrm{s},1}^{H}=1$. 
Then, we prove the sequence of $\{p_{\mathrm{s},1}^{H},p_{\mathrm{s},2}^{H},...,p_{\mathrm{s},n}^{H},p_{\mathrm{s},n+1}^{H},...\}$ is a decreasing sequence. 
For ease of exposition, let us denote by
\begin{equation}
    f(p_\mathrm{s}^{H})=\exp\left(-\frac{\lambda cr^2\eta}{1+A\eta p_\mathrm{s}^{H}}-\frac{\theta r^\alpha}{\rho}\right).
\end{equation}
Note that the function $f(p^{H}_\mathrm{s})$ increases as the increasing $p^{H}_\mathrm{s}$. 
We then use the mathematical indication method, which is
\begin{align}
 p_{\mathrm{s},2}^{H} &= \exp\left(-\frac{\lambda cr^2\eta}{1+A\eta p_{\mathrm{s},1}^{H}}-\frac{\theta r^\alpha}{\rho}\right)\notag\\
 &=  \exp\left(-\frac{\lambda cr^2\eta}{1+A\eta}-\frac{\theta r^\alpha}{\rho}\right)<1=p_{\mathrm{s},1}^{H}, 
\end{align}
and
\begin{align}
 p_{\mathrm{s},3}^{H}&=  \exp\left(-\frac{\lambda cr^2\eta}{1+A\eta p_{\mathrm{s},2}^{H}}-\frac{\theta r^\alpha}{\rho}\right)\notag\\
 &<\exp\left(-\frac{\lambda cr^2\eta}{1+A\eta p_{\mathrm{s},1}^{H}}-\frac{\theta r^\alpha}{\rho}\right)=p_{\mathrm{s},2}^{H}.
\end{align}
Then, suppose $p_{\mathrm{s},n}^{H}  < p_{\mathrm{s},n-1}^{H} $ holds for $n$, we have
\begin{align}
     p_{\mathrm{s},n+1}^{H}&= \exp\left(-\frac{\lambda cr^2\eta}{1+A\eta p_{\mathrm{s},n}^{H}}-\frac{\theta r^\alpha}{\rho}\right)\\
     &<\exp\left(-\frac{\lambda cr^2\eta}{1+A\eta p_{\mathrm{s},n-1}^{H}}-\frac{\theta r^\alpha}{\rho}\right)=p_{\mathrm{s},n}^{H}. 
\end{align}
As such, the sequence of $\{p_{\mathrm{s},1}^{H},p_{\mathrm{s},2}^{H},...,p_{\mathrm{s},n}^{H},p_{\mathrm{s},n+1}^{H},...\}$ is a decrease sequence when $p_{\mathrm{s},1}^{H}=1$. 

With a similar approach, we can prove that the root $p_{\mathrm{s}}^{H}$ is a lower bound of the sequence $\{p_{\mathrm{s},1}^{H},p_{\mathrm{s},2}^{H},...,p_{\mathrm{s},n}^{H},p_{\mathrm{s},n+1}^{H},...\}$. First, we have $p_{\mathrm{s}}^{H}<p_{\mathrm{s},1}^{H}=1$. Then, we have
\begin{align}
p_{\mathrm{s},2}^{H}&= \exp\left(-\frac{\lambda cr^2\eta}{1+A\eta p_{\mathrm{s},1}^{H}}-\frac{\theta r^\alpha}{\rho}\right) \notag \\
&>\exp\left(-\frac{\lambda cr^2\eta}{1+A\eta p_{\mathrm{s}}^{H}}-\frac{\theta r^\alpha}{\rho}\right)=p_{\mathrm{s}}^{H}
\end{align}
 Then, suppose $p_{\mathrm{s},n}^{H}  > p_{\mathrm{s}}^{H} $ holds for $n$, we have
\begin{align}
	p_{\mathrm{s},n+1}^{H}&= \exp\left(-\frac{\lambda cr^2\eta}{1+A\eta p_{\mathrm{s},n}^{H}}-\frac{\theta r^\alpha}{\rho}\right) \notag\\
	&>\exp\left(-\frac{\lambda cr^2\eta}{1+A\eta p_{\mathrm{s}}^{H}}-\frac{\theta r^\alpha}{\rho}\right)=p_{\mathrm{s}}^{H}. 
\end{align}
This completes the proof that $p_\mathrm{s}^{H}$  is a lower bound of the sequence $\{p_{\mathrm{s},1}^{H},p_{\mathrm{s},2}^{H},...,p_{\mathrm{s},n}^{H},p_{\mathrm{s},n+1}^{H},...\}$.

The proof of convergence of $p_\mathrm{s}^{H}$ and its convergence rate is shown as follows:\\
(1) When $\lambda cr^2\eta\leq 4$, the steady-state point $p_\mathrm{s}^{H}$ exists, and we have 
\begin{align}
    &\frac{|p_{\mathrm{s},n+1}^{H}-p_{\mathrm{s},n}^{H}|}{\big{|}p_{\mathrm{s},n}^{H}-p_{\mathrm{s},n-1}^{H}\big{|}} \notag\\ 
    &=\exp{\left(-\frac{\theta r^{\alpha}}{\rho}\right)}\frac{|\exp{\left(-\frac{\lambda cr^2\eta}{1+A\eta p_{\mathrm{s},n}^{H}} \right)}-\exp{\left(-\frac{\lambda cr^2\eta}{1+A\eta p_{\mathrm{s},n-1}^{H}} \right)}|}{|p_{\mathrm{s},n}^{H}-p_{\mathrm{s},n-1}^{H}|}\notag\\
    &\overset{(a)}{=}\frac{|p_{\mathrm{s},n}^{H}-p_{\mathrm{s},n-1}^{H}|\exp\left(-\frac{\lambda cr^2\eta}{1+A\eta \xi}-\frac{\theta r^{\alpha}}{\rho}\right)\frac{\lambda c r^2 A \eta^2}{\left(1+A\eta\xi\right)^2}}{|p_{\mathrm{s},n}^{H}-p_{\mathrm{s},n-1}^{H}|} \notag\\
    &\overset{(b)}{<}\frac{\lambda cr^2 A \eta^2\xi}{\left(1+A\eta\xi\right)^2}=\frac{\lambda cr^2A\eta^2}{A^2\eta^2\xi+\frac{1}{\xi}+2A\eta}\notag\\
    &\overset{(c)}{\leq}\frac{\lambda cr^2\eta}{4}\leq 1.
\end{align}
(2) When $\lambda cr^2\eta>4$ and $A>A_l$, the steady-state point $p_\mathrm{s}^{H}$ exists, we have
\begin{equation}\label{eq:R12}
    \begin{split}
       &\frac{|p_{\mathrm{s},n+1}^{H}-p_{\mathrm{s},n}^{H}|}{\big{|}p_{\mathrm{s},n}^{H}-p_{\mathrm{s},n-1}^{H}\big{|}}\\
        &\overset{(a)}{=}\exp\left(-\frac{\lambda cr^2\eta}{1+A\eta \xi}-\frac{\theta r^{\alpha}}{\rho}\right)\frac{\lambda cr^2 A \eta^2}{\left(1+A\eta\xi\right)^2}\\
        &\overset{(b)}{<}\frac{\lambda cr^2 A \eta^2\xi}{\left(1+A\eta\xi\right)^2}\overset{(d)}{\leq}1,
    \end{split}
\end{equation}
where (a) follows the mean theorem, (b) follows the $\xi>\exp\left(-\frac{\lambda cr^2\eta}{1+A\eta \xi}-\frac{\theta r^{\alpha}}{\rho}\right)$ when $\xi\in \left(p_{\mathrm{s},n+1}^{H},p_{\mathrm{s},n}^{H}\right)$,
the last inequality (c) follows the fundamental inequality, and (d) follows $A>A_l$ and $\xi>p_{\mathrm{s},n+1}^{H}>p_{\mathrm{s}}^{H}>\exp\left(-\frac{1}{\frac{1}{2}+\sqrt{\frac{1}{4}-\frac{1}{\lambda cr^2\eta}}}-\tfrac{\theta r^\alpha}{\rho} \right)$.

The convergence of $p^{L}_\mathrm{s}$ can be shown via similar steps with initial iterative point $p_{\mathrm{s},1}^{L}=\exp\left(-\frac{1}{\frac{1}{2}-\sqrt{\frac{1}{4}-\frac{1}{\lambda cr^2\eta}}}-\tfrac{\theta r^\alpha}{\rho} \right)$, which is upper bound of the $p^{L}_\mathrm{s}$.

Then, the fixed-point iteration in \eqref{eq:define:p:it} always converges with a linear convergence rate.

\section{AoI performance derivation: Theorem 2}\label{proofAoI}
Let us denote $I_n$ as the time interval between two consecutive transmission attempts over the typical link and $J_k$ the waiting time at the destination between successful receptions of the $k$-th and ($k+1$)-th updates, respectively. We have the following relationship between $I_n$ and $J_k$ according to the age threshold-based transmission protocol:
\begin{equation}
	J_k = A+\sum_{n=1}^{N}I_n
\end{equation}
where $N$ is a random variable that represents the number of attempts between two successful transmissions. We further
introduce a variable $L_k$, which represents the area under the AoI evolution curve across the $k$-th successful update, as follows:
\begin{equation}
	L_k = \sum_{j=1}^{J_k}j = \frac{1}{2}J_k(J_k+1).
\end{equation}
Then, we can calculate the mean peak AoI and time average AoI over the typical link as:
\begin{equation} \label{eq:df:AoIp}
	\hat{\Delta}_{0} = \mathbb{E}[J_k],
\end{equation}
and
\begin{equation} \label{eq:df:AoIave}
	\bar{\Delta}_0 = \frac{\mathbb{E}[L_k]}{\mathbb{E}[J_k]} = \frac{1}{2}+\frac{\mathbb{E}[J_k^2]}{2\mathbb{E}[J_k]}.
\end{equation}
Since the transmission success probability $p_\mathrm{s}$ is i.i.d over the time, we can compute $\mathbb{E}[J_k]$ and $\mathbb{E}[J_k^2]$ respectively as the following:
\begin{equation}
	\mathbb{E}[J_k] = A + \sum_{n=1}^{\infty} \mathbb{E}[np_\mathrm{s}(1-p_\mathrm{s})^{n-1}]=A+\frac{\mathbb{E}[I_n]}{p_\mathrm{s}}  ,\label{eq:jk}
\end{equation}
and 
\begin{equation}  \label{eq:jk2}
	\begin{split}
		\mathbb{E}[J_k^2] &= A^2 + 2A  \mathbb{E}\left[ \sum_{n=1}^{N} I_n \right] +  \mathbb{E}\left[ \left(\sum_{n=1}^{N} I_n\right)^2  \right]          \\
		& = A^2 + \frac{2A\mathbb{E}[I_n]+\mathbb{E}[I_n^2]}{p_\mathrm{s}}+\frac{2(1-p_\mathrm{s})\left(\mathbb{E}[I_n]\right)^2}{p_\mathrm{s}^2}.
	\end{split}       
\end{equation}
when a source node is allowed to transmit, it generates new updates with probability $\eta$ independently over time, we have 
\begin{align}
	&\mathbb{E}[I_n] = \frac{1}{\eta},  \label{eq:In}  \\           
	&\mathbb{E}[I_n^2] = \frac{2-\eta}{\eta^2}.  \label{eq:In2}
\end{align}
The age performance is derived by combining \eqref{eq:df:AoIp}, \eqref{eq:df:AoIave}, \eqref{eq:jk}, \eqref{eq:jk2}, \eqref{eq:In} and  \eqref{eq:In2}.

\section{Proof of lemma 1} \label{prooflemma:oppeakq}
By substituting \eqref{eq:define:p} into \eqref{eq:peakAoI}, and take a partial derivative of $\hat{\Delta}_0$ with respect to $\eta$, we have
\begin{equation}
	\frac{\partial{\hat{\Delta}_{0}}}{\partial{\eta}} =-\frac{1}{\eta^2p_\mathrm{s}}+\frac{\lambda cr^2}{\eta p_\mathrm{s}((1+A\eta p_\mathrm{s})^2-\lambda cr^2A\eta^2p_\mathrm{s})}.
\end{equation}
At the two extreme operating points of $\eta$ (i.e., $\eta=0$ and $\eta=1$), we have
\begin{equation}
	\lim_{\eta \to 0}\frac{\partial{\hat{\Delta}_{0}}}{\partial{\eta}} <0,
\end{equation}
and
\begin{equation}
	\lim_{\eta \to 1}\frac{\partial{\hat{\Delta}_{0}}}{\partial{\eta}} = -\frac{1}{p_\mathrm{s}}+\frac{\lambda cr^2}{(1+Ap_\mathrm{s})^2-\lambda cr^2Ap_\mathrm{s}}.
\end{equation}
when $\lambda cr^2 >1+Ap_{\mathrm{s},{*}}$, we have $\underset{{\eta \to 1}}{\lim}\frac{\partial{\hat{\Delta}_{0}}}{\partial{\eta}}>0$, the mean peak AoI can then be optimized when $\eta\in(0,1)$. By combining $\frac{\partial{\hat{\Delta}_{0}}}{\partial{\eta}} =0$ and \eqref{eq:define:p}, the optimal age threshold in mean peak AoI optimization can be obtained as 
\begin{equation}\label{eq:optimal:para:eta:proof}
	\hat{\eta}^{*}_{\vert A} = \dfrac{1}{\lambda cr^2-A\exp\left(-1-\theta r^\alpha\rho^{-1}\right)},
\end{equation}
The optimal mean peak AoI can be obtained by substituting \eqref{eq:optimal:para:eta:proof} into \eqref{eq:peakAoI}.

When $\lambda cr^2 \leq 1+Ap_{\mathrm{s},{*}}$, on the other hand, the optimal age threshold is given by $\eta=1$, and the corresponding optimal mean peak AoI can be obtained by combining $\eta=1$ and \eqref{eq:peakAoI}.

\section{Proof of lemma 2} \label{prooflemma2:peakA}
By substituting \eqref{eq:define:p} into \eqref{eq:peakAoI}, and take a partial derivative of $\hat{\Delta}_0$ with respect to $A$, we have
\begin{equation}
	\frac{\partial{\hat{\Delta}_{0}}}{\partial{A}} =1-\frac{\lambda cr^2\eta}{(1+A\eta p_\mathrm{s})^2-\lambda cr^2A\eta^2p_\mathrm{s}}.
\end{equation}
At the two extreme operating points of $A$ (i.e., $A=0$ and $A\to \infty$), we have
\begin{equation}
	\lim_{A \to 0}\frac{\partial{\hat{\Delta}_{0}}}{\partial{A}} =1-\lambda cr^2 \eta,
\end{equation}
and
\begin{equation}
	\lim_{A \to \infty}\frac{\partial{\hat{\Delta}_{0}}}{\partial{A}} =1,
\end{equation}
when $\lambda cr^2 >\frac{1}{\eta}$, we have $\underset{{A \to 0}}{\lim}\frac{\partial{\hat{\Delta}_{0}}}{\partial{A}}<0$, the mean peak AoI can then be optimized when $A\in(0,\infty)$. By combining $\frac{\partial{\hat{\Delta}_{0}}}{\partial{A}} =0$ and \eqref{eq:define:p}, the optimal age threshold in mean peak AoI optimization can be obtained as 
\begin{equation}\label{eq:optimal:para:A:proof}
	\hat{A}^{*}_{\vert \eta}= \left(\lambda cr^2 -\dfrac{1}{\eta}\right)\exp\left(1+\theta r^\alpha\rho^{-1}\right).
\end{equation}
The optimal mean peak AoI can be obtained by substituting \eqref{eq:optimal:para:A:proof} into \eqref{eq:peakAoI}. 

On the other hand, when $\lambda cr^2 \leq \frac{1}{\eta}$, the optimal age threshold is given by $A=0$, and the corresponding optimal mean peak AoI is obtained by combining $A=0$ and \eqref{eq:peakAoI}.

\section{Proof of Theorem 3}\label{proof:joint:op:peak}
Combining equation \eqref{eq:define:p}, $\frac{\partial{\hat{\Delta}_{0}}}{\partial{\eta}}=0$ and $\frac{\partial{\hat{\Delta}_{0}}}{\partial{A}}=0$, we have
\begin{equation}
	p_\mathrm{s} = \exp\left(-1 - \theta r^\alpha \rho^{-1}\right).
\end{equation}
Then, the optimal mean peak AoI can be achieved when $(\hat{A}^{*},\hat{\eta}^{*})$ satisfy
\begin{equation}
	\hat{A}^{*} = \left(\lambda cr^2- \frac{1}{\hat{\eta}^{*}}\right)\exp\left(1+\theta r^{\alpha}\rho^{-1}\right).
\end{equation}
Due to the fact $A \geq 0$, the optimal parameter pair $(A>0,\eta)$ exists when the condition $\lambda cr^2>\min\{\frac{1}{\eta}\}=1$ is satisfied; Otherwise, the optimal parameter setting is $(A=0,\eta=1)$. The corresponding optimal mean peak AoI can be obtained by substitute $(A,\eta)$ of above two situation in \eqref{eq:peakAoI}. 

\section{Proof of Lemma 3} \label{proof:sole:op:eta}
By substituting \eqref{eq:define:p} into \eqref{eq:averageAoI} and taking a partial derivative of $\bar{\Delta}_0$ with respect to $\eta$, we have
        \begin{footnotesize}
         \begin{equation}
		\frac{\partial{\bar{\Delta}_0}}{\partial{\eta}} =\left(p_{\mathrm{s}}-\frac{\lambda cr^2\eta p_{\mathrm{s}}}{\left(1+A\eta p_{\mathrm{s}}\right)^2-\lambda cr^2A\eta^2p_{\mathrm{s}}}\right)\left(\frac{A}{2(1+A\eta p_{\mathrm{s}})^2}-\frac{1}{\eta^2 p_{\mathrm{s}}^2 }\right).
	\end{equation}   
        \end{footnotesize}
At the two extreme operating points of $\eta$ (i.e., $\eta=0$ and $\eta=1$), we have $\underset{{\eta\to 0}}{\lim}\frac{\partial{\bar{\Delta}}}{\partial{\eta}}\to -\infty$, and 
         \begin{small}
	\begin{align}\label{eq:eta1}
          	&\lim_{\eta\to 1}\frac{\partial{\bar{\Delta}_0}}{\partial{\eta}}\\
       &=\left(p_{\mathrm{s},*}-\frac{\lambda cr^2 p_{\mathrm{s},*}}{\left(1+A p_{\mathrm{s},*}\right)^2-\lambda cr^2Ap_{\mathrm{s},*}}\right)\left(\frac{A}{2(1+Ap_{\mathrm{s},*})^2}-\frac{1}{p_{\mathrm{s},*}^2 }\right).  \notag
	\end{align}
        \end{small}
We notice the last term of \eqref{eq:eta1} satisfies
	\begin{equation}
		\frac{A}{2(1+Ap_{\mathrm{s},*})^2}-\frac{1}{(p_{\mathrm{s},*})^2 }=\frac{Ap_{\mathrm{s},*}^2-2-4Ap_{\mathrm{s},*}-2A^2 p_{\mathrm{s},*}^2}{2(1+Ap_{\mathrm{s},*})^2p_{\mathrm{s},*}^2}<0.
	\end{equation}
Therefore, if the first term of \eqref{eq:eta1} is negative (this is achieved when $\lambda cr^2>1+Ap_{s,*}$), the time average AoI can be optimized by a specific $\eta\in(0,1)$. As such, by solving the equation $\frac{\partial{\bar{\Delta}_0}}{\partial{\eta}}=0$, the optimal update rate is obtained as:
	\begin{equation}\label{eq:optimal:para:eta:proof}
		\bar{\eta}^{*}_{\vert A} = \frac{1}{\lambda cr^2-A\exp\left(-1-\theta r^\alpha\rho^{-1}\right)}.
	\end{equation}
Then, the optimal time-average AoI follows by substituting \eqref{eq:optimal:para:eta:proof} into \eqref{eq:averageAoI}. On the other hand, if $\lambda cr^2 \leq 1+Ap_{s,*}$, the optimal update rate is achieved when $\eta=1$, and the optimal time-average AoI is obtained by combining $\eta=1$ and \eqref{eq:averageAoI}.

\section{Proof of Lemma \ref{Theorem:op:ave:threshold}} \label{prooftheorem5}
By substituting \eqref{eq:define:p} into \eqref{eq:averageAoI}, and taking a partial derivative of $\bar{\Delta}_0$ with respect to $A$, we have
	\begin{equation}
        \begin{split}
		\frac{\partial{\bar{\Delta}_0}}{\partial{A}} &=\frac{A^2\eta^2p_{\mathrm{s}}^2+2A\eta p_{\mathrm{s}}+\eta p_{\mathrm{s}}}{2\left(1+A\eta p_{\mathrm{s}}\right)^2}\\ &-\frac{\lambda cr^2\eta \left(2+4A\eta p_{\mathrm{s}} +A^2\eta^2 p_{\mathrm{s}}^2-A\eta^2p_{\mathrm{s}}^2\right)}{2\left(1+A\eta p_{\mathrm{s}}\right)^2\left(\left(1+A\eta p_{\mathrm{s}}\right)^2-\lambda cr^2A\eta^2p_{\mathrm{s}}\right)}.
        \end{split}
	\end{equation}
The two extreme operating points of the age threshold lead to
$\underset{{A \to 0}}{\lim}\frac{\partial{\bar{\Delta}_0}}{\partial{A}} =\frac{1}{2}\eta p_{\mathrm{s}}^{A=0}-\lambda cr^2 \eta$ and $ \underset{A \to \infty}{\lim} \frac{\partial{\bar{\Delta}_0}}{\partial{A}} = \frac{1}{2}$, respectively.

When $\lambda cr^2\eta>\mathbb{W}_0\left(\frac{\eta}{2}\exp\left(-\theta r^{\alpha}\rho^{-1}\right)\right)$, we have $\underset{{A \to 0}}{\lim}\frac{\partial{\bar{\Delta}_0}}{\partial{A}}<0$. Since $\frac{\partial{\bar{\Delta}_0}}{\partial{A}} $ is continuous in $A$, Intermediate Value Theorem assures that $\frac{\partial{\bar{\Delta}_0}}{\partial{A}} =0$ has solutions in this case. 
As such, by combining $\frac{\partial{\bar{\Delta}_0}}{\partial{A}} =0$ and \eqref{eq:define:p}, the optimal age threshold can be obtained from solving the following equation
\begin{small}
	\begin{equation} \label{eq:op:A:ave:fixpoint}
		(1+A\eta p_{\mathrm{s}})^3-\lambda c r^2\eta(1+A\eta p_{\mathrm{s}})^2+(\eta p_{\mathrm{s}}-1)(1+A\eta p_{\mathrm{s}})-\lambda c r^2\eta=0.
	\end{equation}
\end{small}\eqref{eq:op:AgeThre:ave1} follows by combining \eqref{eq:define:p} and \eqref{eq:op:A:ave:fixpoint}.

The above analysis provides the conditions for the existence of roots. 
Next, we show that \eqref{eq:op:AgeThre:ave1} (resp. \eqref{eq:op:A:ave:fixpoint}) has a unique non-zero root. We denote by $u= 1+A\eta p_{\mathrm{s}}$ and introduce an auxiliary function as follows 
	\begin{equation}
		\begin{split}
			&f(u) = u^3 - \lambda cr^2\eta u^2 \\
			&\quad+\left(\eta\exp\left(\tfrac{\theta r^{\alpha}}{\rho}\right)\exp\left(-\tfrac{\lambda cr^2\eta}{u}\right)-1\right)u- \lambda cr^2\eta.
		\end{split}
	\end{equation}
 Since $f(u)=0$ is equivalent to \eqref{eq:op:AgeThre:ave1}, it can be used to analyze the number of roots.

(1.1) When $\mathbb{W}_0\left(\frac{\eta}{2}\exp\left(-\theta r^{\alpha}\rho^{-1}\right)\right)<\lambda cr^2\eta \leq 1$, the first derivative of $f(u)$ can be expressed as follows:
	\begin{equation}
		\begin{split}
			&\frac{\partial{f(u)}}{\partial{u}}= 3u^2-2\lambda cr^2\eta u-1 \\
			&\quad~~~~+\eta \exp\left(-\tfrac{\theta r^{\alpha}}{\rho}\right)\exp\left(-\tfrac{\lambda cr^2\eta}{u}\right)\left(1+\tfrac{\lambda cr^2\eta}{u}\right).
		\end{split}
	\end{equation}
Because $3u^2-2\lambda cr^2\eta u-1 \geq 0$ for $u\geq 1$ and the last term of $\frac{\partial{f(u)}}{\partial{u}}$ is positive for $u\in[1,\infty)$. We have $\frac{\partial{f(u)}}{\partial{u}}>0$ for $u\in[1,\infty)$. Additionally, since $f(u)$ is monotonically increasing for $u\in[1,\infty)$, $f(u)=0$ has at most one root in this region.

(1.2)  When $1<\lambda cr^2 \eta \leq 3$, the second derivative of $f(u)$ satisfies
	\begin{equation}
		\begin{split}
			&\frac{\partial^2{f(u)}}{\partial{u^2}}= 6u-2\lambda cr^2\eta \\
			&\quad~~ +\eta \exp\left(-\theta r^\alpha \rho^{-1}\right)\exp\left(-\tfrac{\lambda cr^2\eta}{u}\right)\tfrac{(\lambda cr^2\eta)^2}{u^3}>0.
		\end{split}
	\end{equation}
Therefore,  the term $\frac{\partial f(u)}{\partial u }$ increases for $u\in[1,\infty)$. 
We denote by $v=\lambda cr^2\eta$ and let $g(v)= \underset{{u\in[1,\infty)}}{\min} \frac{\partial{f(u)}}{\partial{u}} = \frac{\partial{f(u)}}{\partial{u}}|_{u=1}$, which can be expressed as
	\begin{equation}
		\begin{split}
			&g(v) =2-2v +\eta\exp\left(-\theta r^\alpha \rho^{-1}\right)\exp\left(-v\right)(1+v).
		\end{split}
	\end{equation}
The first derivative of $g(v)$ satisfies $\frac{\partial g(v) }{\partial v }<0$, indicating that $g(v)$ decreases monotonically for $v=\lambda cr^2\eta\in(1,3)$.
Additionally, notice that $g(1) > 0$ and $g(3) < 0$, $g(v)=0$ must have a root in $(1, 3)$, which we denote by $\psi$. 
Consequently, if 
if $\lambda cr^2\in(1,\psi]$, $f(u)$ increases for $u\in(1,\infty)$; if $\lambda cr^2\in(\psi,3]$, $f(u)$ decreases first, and then increases with $u$. Moreover, since $f(1)<0$ and  $\underset{u \to \infty}{\lim}f(u)>0$, $f(u)$ has a unique root in this case.

(1.3) When $\lambda cr^2 \eta >3$,   the third derivative of $f(u)$ satisfy\begin{equation}
		\begin{split}
			&\frac{\partial^3{f(u)}}{\partial{u^3}}= 6+\eta \exp\left(-\theta r^\alpha \rho^{-1}\right)\times\\
			&~~~~\exp\left(-\frac{\lambda cr^2\eta}{u}\right)\frac{(\lambda cr^2\eta)^2(\lambda c r^2\eta -3u)}{u^5}\\
			&>6+\frac{(\lambda cr^2\eta)^3}{u^3}\exp\left(-\frac{\lambda cr^2\eta}{u}\right)\left(\frac{1-u}{u^2}\right)\\
			&>6-\frac{(\lambda cr^2\eta)^3}{u^3}\exp\left(-\frac{\lambda cr^2\eta}{u}\right)>0.
		\end{split}
	\end{equation}
Then, the second derivative $\frac{\partial^2{f(u)}}{\partial{u^2}}$ increases for $u \in[1,\infty)$. 

(a) If $\frac{\partial^2{f(u)}}{\partial{u^2}}|_{u=1}<0$, $\frac{\partial{f(u)}}{\partial{u}}$ decreases first, then increases for $u\in[1,\infty)$, and   $\frac{\partial{f(u)}}{\partial{u}}|_{u=1}<0$, $\underset{u \to \infty}{\lim}\frac{\partial{f(u)}}{\partial{u}}>0$. Then, $f(u)$ decreases first and then increases versus $u$. Since $f(1)<0$ and  $\underset{u \to \infty}{\lim}f(u)>0$, $f(u)$ has a root at most.

(b) If $\frac{\partial^2{f(u)}}{\partial{u^2}}|_{u=1} \geq 0$, $\frac{\partial{f(u)}}{\partial{u}}$ monotonically increases for $u\in[1,\infty)$, and   $\frac{\partial{f(u)}}{\partial{u}}|_{u=1}<0$, $\underset{u \to \infty}{\lim}\frac{\partial{f(u)}}{\partial{u}}>0$. Then, $f(u)$ decreases first and then increases versus $u$. Since $f(1)<0$ and  $\underset{u \to \infty}{\lim}f(u)>0$, $f(u)=0$ has a single root.

The discussions in (1.1), (1.2), and (1.3) confirm that \eqref{eq:op:A:ave:fixpoint} has a unique root, and the optimal age threshold $A_{1}$ is thus also unique. The results further indicate that when $\lambda cr^2\eta > \mathbb{W}_0\left(\frac{\eta}{2}\exp\left(-\theta r^{\alpha}\rho^{-1}\right)\right)$, $\bar{\Delta}_{0}$ decreases first, then increases with $A$, the solution of \eqref{eq:op:AgeThre:ave1} is the global minimum.

When $\lambda cr^2\eta \leq \mathbb{W}_0\left(\frac{\eta}{2}\exp\left(-\theta r^{\alpha}\rho^{-1}\right)\right)$, $\bar{\Delta}_0$ monotonically increases for $A\in[0,+\infty)$. The optimal age threshold is thus given by $A=0$, and the corresponding optimal time-average AoI can be obtained by combining $A=0$ and \eqref{eq:averageAoI}.

\section{Proof of Theorem \ref{Theorem:op:ave:joint}} \label{proof:op:ave:joint} 
We denote by $\bar{\Delta}_0^{*} = \underset{\{\eta\}}{\min} ~\bar{\Delta}_0^{A=\tilde{A}^{*}_{\vert \eta}}$. 
On the one hand, for $\lambda cr^2\leq\frac{1}{\eta}\mathbb{W}_0\left(\frac{\eta}{2}\exp\left(- \frac{ \theta r^{\alpha} }{ \rho } \right)\right)$, we have  
\begin{small}
	\begin{equation}
		\bar{\Delta}_0^{A=A^{*}_{\vert \eta}} = \frac{1}{\eta}\exp\left( \lambda cr^2\eta+\theta r^\alpha \rho^{-1}\right).
	\end{equation}
\end{small}Then, we take a derivative of $\bar{\Delta}_0^{A=A^{*}_{\vert \eta}}$ with respect to $\eta$, which gives
\begin{small}
	\begin{equation}
		\tfrac{\partial{\bar{\Delta}_0^{A=A^{*}_{\vert \eta}}}}{\partial{\eta}} =\frac{\lambda cr^2\eta-1}{\eta^2} \exp\left( \lambda cr^2\eta+\theta r^\alpha \rho^{-1}\right).
	\end{equation}
\end{small}Following this result, we note that if 
\begin{equation}
 \lambda cr^2<\frac{1}{\eta}\mathbb{W}_0\left(\frac{\eta}{2}\exp\left(- \frac{ \theta r^{\alpha} }{ \rho } \right)\right),  
\end{equation}
Due to the fact, 
\begin{equation}
   \frac{1}{\eta}\mathbb{W}_0\left(\frac{\eta}{2}\exp\left(- \frac{ \theta r^{\alpha} }{ \rho } \right)\right)<\frac{1}{\eta}\mathbb{W}_0\left(\frac{1}{2}\right)<\frac{1}{\eta},
\end{equation}
there is $\frac{\partial{\bar{\Delta}_0^{A=A^{*}_{\vert \eta}}}}{\partial{\eta}}<0$, namely $\bar{\Delta}_0^{A=A^{*}_{\vert \eta}} $ decreases with the update rate $\eta$ in this region.

On the other hand, when $\lambda cr^2>\frac{1}{\eta}\mathbb{W}_0\left(\frac{\eta}{2}\exp\left(- \frac{ \theta r^{\alpha} }{ \rho } \right)\right)$,
$\bar{\Delta}_0^{A=\bar{A}^{*}_{\vert \eta}}$ can be explicitly approximated by \eqref{eq:asymptoticAAoI}, and we have
\begin{small}
	\begin{equation}
		\tfrac{\partial{\bar{\Delta}_0^{A=\tilde{A}^{*}_{\vert \eta}}}}{\partial{\eta}} = \tfrac{-3\left((\Omega^{\frac{1}{3}})^{'} G(\eta) - \Omega^{\frac{1}{3}}G^{'}(\eta)\right)}{2(G(\eta))^2}\\
		-\tfrac{3\Omega^{\frac{1}{3}}\exp\left(\frac{3\lambda cr^2\eta\Omega^{\frac{1}{3}}}{G(\eta)} +\frac{ \theta r^\alpha }{ \rho } \right) }{\eta ^2 G(\eta)},
	\end{equation}
\end{small}where $G(\eta) = \Omega^{\frac{2}{3}}+\lambda cr^2\eta\Omega^{\frac{1}{3}}+(\lambda cr^2\eta)^2+3$ and $\Omega$ is given in \eqref{eq:omega}. Since the exponential term is greater than one, then 
\begin{small}
	\begin{align}\label{eq:51}
		&	\dfrac{\partial{\bar{\Delta}_0^{A=\tilde{A}^{*}_{\vert \eta}}}}{\partial{\eta}}\leq -\frac{3\eta((\Omega^{\frac{1}{3}})^{'}\eta G(\eta) - \Omega^{\frac{1}{3}}\eta G^{'}(\eta))}{2\eta^2(G(\eta))^2}-  \frac{6\eta\Omega^{\frac{1}{3}}G(\eta)}{2\eta ^2 (G(\eta))^2} \notag\\
		&\quad~\overset{(a)}{=}\frac{3\left(G^{'}(\eta)\eta\Omega^{\frac{1}{3}}-G(\eta)(\Omega^{\frac{1}{3}}+\eta (\Omega^{\frac{1}{3}})^{'})\right)}{2\eta(G(\eta))^2}-  \frac{3\Omega^{\frac{1}{3}}}{2\eta G(\eta)}\overset{(b)}{<}0,
	\end{align}
\end{small}where ($a$) follows algebraic manipulations. We further clarify the second inequality ($b$) in the following. 

Since $G(\eta)$, $\Omega$, and $\eta$ are positive, we only need to show that $G^{'}(\eta)\eta\Omega^{\frac{1}{3}}-G(\eta)(\Omega^{\frac{1}{3}}+\eta (\Omega^{\frac{1}{3}})^{'})<0$. We expand this term as follows:
\begin{small}
	\begin{equation}\label{eq:51:proof3}
		\begin{split}
			&G^{'}(\eta)\eta\Omega^{\frac{1}{3}}-G(\eta)(\Omega^{\frac{1}{3}}+\eta (\Omega^{\frac{1}{3}})^{'}) = -\eta^2(\lambda cr^2-1)^2\Omega^{\frac{1}{3}}\\
			&\quad\quad~~+\frac{\Omega^{'}\eta}{3\Omega^{\frac{2}{3}}} \left(\Omega^{\frac{2}{3}}-(\lambda cr^2\eta)^2 -3\right)-\Omega-(3-\eta^2)\Omega^{\frac{1}{3}}.
		\end{split}
	\end{equation}
\end{small}In \eqref{eq:51:proof3}, the sign of terms in the second line requires further clarification. The $\Omega^{'}\eta$ in this term can be expressed as
\begin{small}
	\begin{equation}
		\Omega^{'}\eta = \Omega+ 2(\lambda cr^2\eta)^3+3\sqrt{3}\frac{(\lambda cr^2\eta)^4+1}{\sqrt{(\lambda cr^2\eta)^4 + 11(\lambda cr^2\eta)^2 -1}},
	\end{equation}
\end{small}which satisfy $	\Omega^{'}\eta<3\Omega$. Then, when $\Omega^{\frac{2}{3}}-(\lambda cr^2\eta)^2 -3>0$, the terms in second line satisfy the following inequality
\begin{small}
	\begin{equation}
		\begin{split}
			&\quad~~\frac{\Omega^{'}\eta}{3\Omega^{\frac{2}{3}}} \left(\Omega^{\frac{2}{3}}-(\lambda cr^2)^2\eta^2 -3\right)-\Omega-(3-\eta^2)(\Omega^{\frac{1}{3}})\\
			&~<\Omega^{\frac{1}{3}}\left(\Omega^{\frac{2}{3}}-(\lambda cr^2)^2\eta^2 -3\right)-\Omega-(3-\eta^2)(\Omega^{\frac{1}{3}})<0.
		\end{split}
	\end{equation}
\end{small}Then, $G^{'}(\eta)\eta\Omega^{\frac{1}{3}}-G(\eta)(\Omega^{\frac{1}{3}}+\eta (\Omega^{\frac{1}{3}})^{'})<0$.
On the other hand, when $ \Omega^{\frac{2}{3}}-(\lambda cr^2\eta)^2-3 < 0$, due to $	\Omega^{'}\eta>0$,  the condition  $G^{'}(\eta)\eta\Omega^{\frac{1}{3}}-G(\eta)(\Omega^{\frac{1}{3}}+\eta (\Omega^{\frac{1}{3}})^{'})<0$ is still satisfied. Therefore, $\frac{\partial{\bar{\Delta}_0^{\tilde{A}^{*}_{\vert \eta}}}}{\partial{\eta}}<0$,
which means that $\bar{\Delta}_0^{A=A^{*}_{\vert \eta}} $ decreases monotonically with update rate $\eta$ in this region.

Therefore, $\bar{\Delta}_0^{*} = \underset{\{\eta\}}{\min} ~\bar{\Delta}_0^{A=\tilde{A}^{*}_{\vert \eta}}$ when $\eta =1$, and Theorem \ref{Theorem:op:ave:joint} can be obtained by combining Lemma \ref{Theorem:op:ave:threshold}, $\eta=1$ and \eqref{eq:SubOptAgeThrld}.

\bibliography{reference}

\begin{thebibliography}{10}

\bibitem{pappas2023age}
N.~Pappas, M.~A. Abd-Elmagid, B.~Zhou, W.~Saad, and H.~S. Dhillon, {\em Age of
  Information: Foundations and Applications}.
\newblock Cambridge University Press, 2023.

\bibitem{AoIinfocom2012}
S.~Kaul, R.~Yates, and M.~Gruteser, ``Real-time status: How often should one
  update?,'' in {\em Proc. IEEE INFOCOM, Orlando, FL, USA}, pp.~2731--2735,
  Mar. 2012.

\bibitem{6195689}
R.~D. Yates and S.~K. Kaul, ``The age of information: Real-time status updating
  by multiple sources,'' {\em IEEE Trans. Inf. Theory}, vol.~65, no.~3,
  pp.~1807--1827, Sep. 2019.

\bibitem{8006544}
R.~D. {Yates} and S.~K. {Kaul}, ``Status updates over unreliable multiaccess
  channels,'' in {\em Proc. IEEE Int. Symp. Inf. Theory (ISIT), Aachen,
  Germany}, pp.~331--335, Jun. 2017.

\bibitem{10114593}
W.~Zhan, D.~Wu, X.~Sun, Z.~Guo, P.~Liu, and J.~Liu, ``Peak age of information
  optimization of slotted {Aloha}: {FCFS} versus {LCFS},'' {\em IEEE Trans.
  Netw. Sci. Eng.}, pp.~1--13, May 2023.

\bibitem{9181539}
B.~Yu, Y.~Cai, and D.~Wu, ``Joint access control and resource allocation for
  short-packet-based mmtc in status update systems,'' {\em IEEE J. Sel. Areas
  Commun.}, vol.~39, no.~3, pp.~851--865, Aug. 2021.

\bibitem{TwoUserMA}
M.~Salimnejad and N.~Pappas, ``On the age of information in a two-user multiple
  access setup,'' {\em Entropy}, vol.~24, no.~4, pp.~542--563, Apr. 2022.

\bibitem{IRSA}
A.~Munari, ``Modern random access: An age of information perspective on
  irregular repetition slotted {ALOHA},'' {\em IEEE Trans. Commun.}, vol.~69,
  no.~6, pp.~3572--3585, Feb. 2021.

\bibitem{ICC2015}
F.~Vázquez-Gallego, J.~Alonso-Zarate, and L.~Alonso, ``Reservation dynamic
  frame slotted-{ALOHA} for wireless {M2M} networks with energy harvesting,''
  in {\em Proc. IEEE ICC, London, UK}, pp.~5985--5991, Jun. 2015.

\bibitem{ThresholdInfocomWS}
H.~{Chen}, Y.~{Gu}, and S.~C. {Liew}, ``Age-of-information dependent random
  access for massive {IoT} networks,'' in {\em Proc. IEEE INFOCOM Workshop,
  Toronto, Canada}, pp.~930--935, Jul. 2020.

\bibitem{yavascan2020analysis}
O.~T. Yavascan and E.~Uysal, ``Analysis of slotted {ALOHA} with an age
  threshold,'' {\em IEEE J. Sel. Areas Commun.}, vol.~39, no.~5,
  pp.~1456--1470, May 2021.

\bibitem{ThresholdISIT}
X.~Chen, K.~Gatsis, H.~Hassani, and S.~S. Bidokhti, ``Age of information in
  random access channels,'' {\em IEEE Trans. Inf. Theory}, vol.~68, no.~10,
  pp.~6548--6568, Jun. 2022.

\bibitem{ThresholdWhittleIndex}
J.~{Sun}, Z.~{Jiang}, B.~{Krishnamachari}, S.~{Zhou}, and Z.~{Niu},
  ``Closed-form {Whittle}’s index-enabled random access for timely status
  update,'' {\em IEEE Trans. Commun.}, vol.~68, no.~3, pp.~1538--1551, Mar.
  2020.

\bibitem{MiSTA}
M.~Ahmetoglu, O.~T. Yavascan, and E.~Uysal, ``Mista: An age-optimized slotted
  {ALOHA} protocol,'' {\em IEEE Internet of Things J.}, vol.~9, no.~17,
  pp.~15484--15496, May 2022.

\bibitem{TCOM2024AFSA}
M.~Moradian, A.~Dadlani, A.~Khonsari, and H.~Tabassum, ``Age-aware dynamic
  frame slotted {ALOHA} for machine-type communications,'' {\em IEEE Trans.
  Commun.}, vol.~72, no.~5, pp.~2639--2654, May 2024.

\bibitem{yang2023analysis}
H.~H. Yang, N.~Pappas, T.~Q.~S. Quek, and M.~Haenggi, ``Analysis of the age of
  information in age-threshold slotted {ALOHA},'' in {\em Int. Symposium
  Modeling and Optimization in Mobile, Ad Hoc, and Wireless Networks (WiOpt),
  Singapore}, pp.~366--373, Aug. 2023.

\bibitem{TITCollisionModel}
J.~Massey and P.~Mathys, ``The collision channel without feedback,'' {\em IEEE
  Trans. Inf. Theory}, vol.~31, no.~2, pp.~192--204, Mar. 1985.

\bibitem{ZhongyiSG}
Y.~{Hu}, Y.~{Zhong}, and W.~{Zhang}, ``Age of information in {Poisson}
  networks,'' in {\em Proc. Int. Conf. Wireless Commun. Signal Process. (WCSP)
  Hangzhou, China}, pp.~1--6, Oct. 2018.

\bibitem{mankar2020spatial}
P.~D. Mankar, M.~A. Abd-Elmagid, and H.~S. Dhillon, ``Spatial distribution of
  the mean peak age of information in wireless networks,'' {\em IEEE Trans.
  Wireless Commun.}, vol.~20, no.~7, pp.~4465--4479, Feb. 2021.

\bibitem{TimeEventAoI}
M.~{Emara}, H.~{Elsawy}, and G.~{Bauch}, ``A spatiotemporal model for peak
  {AoI} in uplink {IoT} networks: Time versus event-triggered traffic,'' {\em
  IEEE Internet of Things J.}, vol.~7, no.~8, pp.~6762--6777, Aug. 2020.

\bibitem{yangUnderstanding}
H.~H. Yang, C.~Xu, X.~Wang, D.~Feng, and T.~Q.~S. Quek, ``Understanding age of
  information in large-scale wireless networks,'' {\em IEEE Trans. Wireless
  Commun.}, vol.~20, no.~5, pp.~3196--3210, May 2021.

\bibitem{LCFSyang}
H.~H. Yang, A.~Arafa, T.~Q.~S. Quek, and H.~V. Poor, ``Spatiotemporal analysis
  for age of information in random access networks under last-come first-serve
  with replacement protocol,'' {\em IEEE Trans. Wireless Commun.}, vol.~21,
  no.~4, pp.~2813--2829, Sep. 2022.

\bibitem{SGThroughputAoI}
P.~D. Mankar, Z.~Chen, M.~A. Abd-Elmagid, N.~Pappas, and H.~S. Dhillon,
  ``Throughput and age of information in a cellular-based {IoT} network,'' {\em
  IEEE Trans. Wireless Commun.}, vol.~20, no.~12, pp.~8248--8263, Jun. 2021.

\bibitem{AoIopSun}
X.~Sun, F.~Zhao, H.~H. Yang, W.~Zhan, X.~Wang, and T.~Q.~S. Quek, ``Optimizing
  age of information in random-access {Poisson} networks,'' {\em IEEE Internet
  of Things J.}, vol.~9, no.~9, pp.~6816--6829, Sep. 2022.

\bibitem{VTC}
F.~Zhao, X.~Sun, W.~Zhan, X.~Wang, and X.~Chen, ``Information freshness in
  random-access {Poisson} network: Average {AoI} versus peak {AoI},'' in {\em
  Proc. IEEE VTC2022-Fall, London, United Kingdom}, pp.~1--6, Sep. 2022.

\bibitem{energyAoItradeoff}
F.~Zhao, X.~Sun, W.~Zhan, X.~Wang, J.~Gong, and X.~Chen, ``Age-energy tradeoff
  in random-access {Poisson} networks,'' {\em IEEE Trans. Green Commun. Netw.},
  vol.~6, no.~4, pp.~2055--2072, Dec. 2022.

\bibitem{FSA}
Z.~Yue, H.~H. Yang, M.~Zhang, and N.~Pappas, ``Age of information under frame
  slotted {ALOHA}-based status updating protocol,'' {\em IEEE J. Sel. Areas
  Commun.}, vol.~41, no.~7, pp.~2071--2089, May 2023.

\bibitem{TMCSG}
H.~H. Yang, A.~Arafa, T.~Q.~S. Quek, and H.~V. Poor, ``Optimizing information
  freshness in wireless networks: A stochastic geometry approach,'' {\em IEEE
  Trans. Mobile Comput.}, vol.~20, no.~6, pp.~2269--2280, Jun. 2021.

\bibitem{TMClocal}
H.~H. Yang, M.~Song, C.~Xu, X.~Wang, and T.~Q.~S. Quek, ``Locally adaptive
  status updating for optimizing age of information in {Poisson} networks,''
  {\em IEEE Trans. Mobile Comput.}, vol.~22, no.~12, pp.~7343--7354, Dec. 2023.

\bibitem{TMCpowercontrol}
M.~Song, H.~H. Yang, H.~Shan, J.~Lee, and T.~Q.~S. Quek, ``Age of information
  in wireless networks: Spatiotemporal analysis and locally adaptive power
  control,'' {\em IEEE Trans. Mobile Comput.}, vol.~22, no.~6, pp.~3123--3136,
  Dec. 2023.

\bibitem{SGbookBac}
F.~Baccelli and B.~Blaszczyszyn, {\em Stochastic Geometry and Wireless
  Networks: Volume I Theory}.
\newblock Boston, MA, USA: Now Foundations and Trends, 2009.

\bibitem{SGbookMartin}
M.~Haenggi, {\em Stochastic Geometry for Wireless Networks}.
\newblock Cambridge University Press, 2012.

\bibitem{TITcapacity}
P.~Gupta and P.~Kumar, ``The capacity of wireless networks,'' {\em IEEE Trans.
  Inf. Theory}, vol.~46, no.~2, pp.~388--404, Mar. 2000.

\bibitem{MartinRandomDistance}
M.~Haenggi, ``On distances in uniformly random networks,'' {\em IEEE Trans.
  Inf. Theory}, vol.~51, no.~10, pp.~3584--3586, Oct. 2005.

\bibitem{MartinReyleignDistance}
S.~S. Kalamkar and M.~Haenggi, ``The spatial outage capacity of wireless
  networks,'' {\em IEEE Trans. Wireless Commun.}, vol.~17, no.~6,
  pp.~3709--3722, Jun. 2018.

\bibitem{TWCUplinkSyn}
M.~Hyder and K.~Mahata, ``Zadoff–chu sequence design for random access
  initial uplink synchronization in lte-like systems,'' {\em IEEE Trans.
  Wireless Commun.}, vol.~16, no.~1, pp.~503--511, Jan. 2017.

\bibitem{IoTJUplinkSyn}
H.~Chougrani, S.~Kisseleff, W.~A. Martins, and S.~Chatzinotas, ``Nb-iot random
  access for nonterrestrial networks: Preamble detection and uplink
  synchronization,'' {\em IEEE Internet of Things J.}, vol.~9, no.~16,
  pp.~14913--14927, Aug. 2022.

\bibitem{SunyinTIT}
Y.~Sun, E.~Uysal-Biyikoglu, R.~D. Yates, C.~E. Koksal, and N.~B. Shroff,
  ``Update or wait: How to keep your data fresh,'' {\em IEEE Trans. Inf.
  Theory}, vol.~63, no.~11, pp.~7492--7508, Nov. 2017.

\bibitem{MartinHM}
K.~Stamatiou and M.~Haenggi, ``Random-access {Poisson} networks: Stability and
  delay,'' {\em IEEE Commun. Lett.}, vol.~14, no.~11, pp.~1035--1037, Nov.
  2010.

\bibitem{MartinMoveLocaldelay}
Z.~Gong and M.~Haenggi, ``The local delay in mobile {Poisson} networks,'' {\em
  IEEE Trans. Wireless Commun.}, vol.~12, no.~9, pp.~4766--4777, Sep. 2013.

\bibitem{UnifiedSINRyang}
H.~H. Yang, T.~Q.~S. Quek, and H.~Vincent~Poor, ``A unified framework for sinr
  analysis in {Poisson} networks with traffic dynamics,'' {\em IEEE Trans.
  Commun.}, vol.~69, no.~1, pp.~326--339, Jan. 2021.

\bibitem{7345601}
M.~Haenggi, ``The meta distribution of the {SIR} in {Poisson} bipolar and
  cellular networks,'' {\em IEEE Trans. Wireless Commun.}, vol.~15, no.~4,
  pp.~2577--2589, Dec. 2016.

\bibitem{localDelayMartin}
M.~Haenggi, ``The local delay in {Poisson} networks,'' {\em IEEE Trans. Inf.
  Theory}, vol.~59, no.~3, pp.~1788--1802, Nov. 2013.

\bibitem{DaiLinALOHACSMA}
L.~Dai, ``Toward a coherent theory of {CSMA} and {ALOHA},'' {\em IEEE Trans.
  Wireless Commun.}, vol.~12, no.~7, pp.~3428--3444, Jun. 2013.

\bibitem{BistableALOHA}
A.~Carleial and M.~Hellman, ``Bistable behavior of {ALOHA}-type systems,'' {\em
  IEEE Trans. Commun.}, vol.~23, no.~4, pp.~401--410, Apr. 1975.

\bibitem{ZhanM2M}
W.~Zhan and L.~Dai, ``Massive random access of machine-to-machine
  communications in lte networks: Modeling and throughput optimization,'' {\em
  IEEE Trans. Wireless Commun.}, vol.~17, no.~4, pp.~2771--2785, Feb. 2018.

\bibitem{DAILINStableALOHA}
L.~Dai, ``A theoretical framework for random access: Stability regions and
  transmission control,'' {\em IEEE/ACM Trans. Networking}, vol.~30, no.~5,
  pp.~2173--2200, Oct. 2022.

\bibitem{lambertW}
{R.~M.~Corless, G.~H.~Gonnet, D.~E.~G.~Hare, D.~J.~Jeffrey and D.~H.~Knuth},
  ``On the {Lambert} {W} function,'' {\em Adv. Comput. Math.}, vol.~5,
  pp.~329--359, Dec. 1996.

\end{thebibliography}

\end{document}